\documentclass[journal]{IEEEtran}

\usepackage{xcolor}
\usepackage{amsmath,amssymb,amsfonts}
\usepackage{mathtools}
\DeclarePairedDelimiter{\ceil}{\lceil}{\rceil}
\usepackage{algorithmic}
\usepackage{graphicx}
\usepackage{subfigure}
\usepackage{textcomp}
\usepackage{xcolor}
\usepackage{comment}
\usepackage{enumerate}

\usepackage{multicol}
\usepackage[noadjust]{cite}

\ifCLASSINFOpdf
\else
\fi
\begin{document}
%
\title{Battery-Less LoRaWAN Communications using Energy Harvesting: Modeling and Characterization}
%
%
%


%
    
    
    

\author{Carmen Delgado, Jos\'e Mar\'ia Sanz, Chris Blondia, and Jeroen Famaey
	\thanks{Carmen Delgado, Chris Blondia and Jeroen Famaey are with IDLab, University of Antwerp -- imec, Antwerp, Belgium, E-mail: carmen.delgado@uantwerpen.be}
	\thanks{Jos\'e Mar\'ia Sanz is with Kintech Engineering, Zaragoza, Spain}
\thanks{Copyright (c) 2020 IEEE. Personal use of this material is permitted. However, permission to use this material for any other purposes must be obtained from the IEEE by sending a request to pubs-permissions@ieee.org.}}

\markboth{Submitted to IEEE Internet of Things Journal}%
{Shell \MakeLowercase{\textit{et al.}}: Bare Demo of IEEEtran.cls for IEEE Journals}
%



\maketitle

\begin{abstract}
Billions of IoT devices are deployed worldwide and batteries are their main power source. However, these batteries are bulky, short-lived and full of hazardous chemicals that damage our environment. 
Relying on batteries is not a sustainable solution for the future IoT. As an alternative, battery-less devices 
run on long-lived capacitors charged using energy harvesters. 
The small energy storage capacity of capacitors  
results in an intermittent on-off behaviour. LoRaWAN is a popular Low Power Wide Area Network technology used in many IoT devices and can be used in these new scenarios. In this work, we present a Markov model to characterize the performance of battery-less LoRaWAN devices for uplink and downlink transmissions and we evaluate their performance in terms of the parameters that define the model (i.e., device configuration, application behaviour and environmental conditions). Results show that LoRaWAN battery-less communications are feasible if choosing the proper configuration (i.e., capacitor size, turn-on voltage threshold) for different application behaviour (i.e., transmission interval, UL/DL packet sizes) and environmental conditions (i.e., energy harvesting rate). 
Since downlink in the second reception window highly affects the performance, only small DL packet sizes should be considered for these devices. Besides, a 47 mF capacitor can support 1 Byte $SF7$ transmissions every 60 s at an energy harvesting rate of 1 mW. However, if no DL is expected, a 4.7 mF capacitor could support 1 Byte $SF7$ transmissions every 9~s. 

\end{abstract}


%
\IEEEpeerreviewmaketitle

\maketitle

\section{Introduction}\label{sec:intro}

\IEEEPARstart{I}{n} 
the Internet of Things (IoT) vision, tens of billions of interconnected devices cooperate to sense, actuate, locate and communicate with each other over the Internet with the aim of supporting and improving daily life. Recent advancements in ultra-low power communications technologies are driving the transformation of everyday objects into an information source connected to the Internet.  Usually, these devices are equipped with a battery, a radio chip, a microcontroller unit (MCU) and one or more sensors and/or actuators.

Low-Power Wide-Area Networks (LPWANs) are a new set of radio technologies that are designed to support the needs of IoT deployments by combining low energy consumption with long range communications \cite{Raza2017}. LoRaWAN \cite{LoraWan} is an LPWAN protocol that builds on top of the LoRa modulation and radio technology,  
uses sub-GHz unlicensed spectrum and enables long-range transmissions (more than 10 km in rural areas) at low power consumption~\cite{Sanchez2018}. 

However, from the beginning, battery lifetime limitations have been one of the main problems to realise the vision of a global IoT. Sometimes, devices are located in hard-to-reach areas, and battery replacement or maintenance is not only costly, but also dangerous. 
In other situations, small devices are needed, but batteries are too bulky. In general terms, it is known that batteries are expensive, bulky and hazardous; they are sensitive to temperature, short-lived, and therefore incompatible with a sustainable IoT~\cite{Hester2017}. 
Moreover, batteries suffer more degradation in combination with current peaks. IoT devices usually spend most of their time in sleep mode until they wake up to perform their tasks, which normally include the transmission and/or reception of data. These two tasks consume much more energy than the sleep mode. As such, current peaks often occur during transmission and reception in IoT devices. For this reason, batteries will degrade faster than expected.
Besides, battery maintenance and replacement mean an increase in the operational costs of IoT deployments, which limits its potential impact in terms of upscaling the number of devices, its deployment in hard-to-reach areas and also the potential lifetime of networks since batteries deplete over time and rechargeable capacity degrades.

Since the number of IoT devices will continue to increase in the coming years, it seems clear that the use of batteries should be reconsidered. 
To alleviate the IoT's battery problem, battery-less IoT devices are a promising solution. 
Battery-less IoT devices are smaller, live longer, are more environmentally friendly, and cheaper to maintain.
This makes them especially suitable for applications in hard-to-reach locations (e.g., intra-body health monitoring, remote-area sensing) and large-scale deployments (e.g., dense building automation networks, smart cities).

In fact, the three main application areas where battery-less devices will benefit are: (i) hard-to-reach or material-embedded devices, (ii) massive-scale IoT networks, and (iii) long-lifetime deploy-and-forget devices. First,
 devices embedded into materials or deployed in remote (e.g., mountainous regions), dangerous (e.g., sewage pipes) or other hard-to-reach locations are hard, dangerous and costly to access for battery maintenance. Such devices often cannot be accessed for battery replacement without damaging the materials in which they are embedded. Additionally, chemical batteries may pose a danger if implanted into living tissue (e.g., skin implants). Second, although massively deployed devices may be easily accessible, the scale of the deployment causes battery maintenance to be exceedingly time-consuming, complex and costly. 
This is for example relevant for fine-grained environmental monitoring (e.g., air quality, occupancy) in buildings or cities, or for tracking goods in large logistics warehouses.
Third, even if in some cases battery replacement may be economically feasible, it might be unwanted due to other reasons, such as ease-of-use or maintenance-free time requirement. For example, it may not be feasible to expect the user to manually charge their device or replace batteries (e.g., wearables for the elderly). In certain scenarios, devices may also have a minimal required maintenance-free lifetime.

\begin{figure}[t]
\centerline{\includegraphics[width=1\columnwidth]{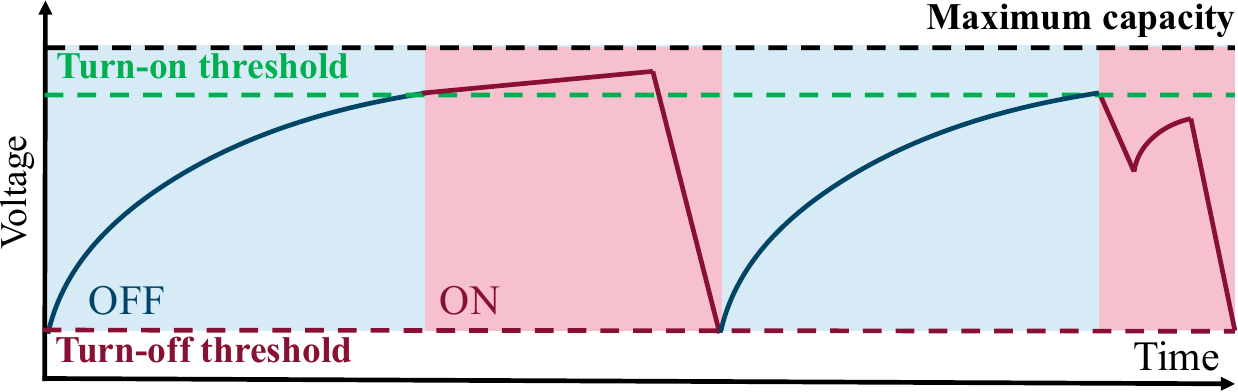}}
\caption{Turn-off and turn-on threshold of battery-less intermittent behaviour}
\label{fig:behaviour}
\end{figure}

In contrast to their battery-powered counterparts, battery-less
devices need to harvest energy from their environment and store it in tiny capacitors. Such capacitors easily last more than ten years \cite{Spanik2014} thanks to the fact that they can handle a much larger number of recharge cycles than rechargeable batteries. Moreover, they are cheaper to produce and easy to recycle, thus better for the environment. However, this new paradigm faces some difficulties: energy harvesting is inconsistent, energy storage is scarce, power failures are inevitable, and execution is intermittent \cite{Hester2017}. This intermittency (c.f Figure~\ref{fig:behaviour}) causes the device to turn on and off frequently, as it swiftly depletes the energy stored in the capacitor. This results in a power failure when the capacity voltage drops below the turn-off threshold. When the device harvests enough energy, it will turn on again when the turn-on voltage threshold is reached, which is a configurable parameter.

In our previous work \cite{Delgado2019} we presented a system model for a battery-less LoRaWAN Class A device where only uplink transmissions were considered and that was evaluated through an event-based simulation system. 
The relatively long runtime of the simulator to determine average performance makes it ill-suited for use in real-time optimization of the system parameters.
For this reason, we extend the simulator to include downlink and we present a Markov Chain model that estimates the average performance in a more efficient way. We also better characterize the device performance giving more in-depth results in terms of reliability for different device configurations (e.g., capacitor size, turn-on voltage threshold), environmental conditions (e.g., energy harvesting rate) and application requirements (e.g., transmission interval, packet size, downlink packet size).

The remainder of this paper is organized as follows. Section~\ref{sec:relatedwork} provides an overview of the related work in the field of battery-less devices, energy harvesting  and LoRaWAN communications. Section~\ref{sec:model} describes the proposed battery-less LoRaWAN device model. In Section~\ref{sec:MC model}, we introduce the Markov Chain model.
The evaluation of the battery-less IoT device model is provided in Section~\ref{sec:modelevaluation}.
In Section~\ref{sec:accuracy}, the Markov Chain model is compared with the simulator to evaluate its accuracy, and a deep insight in the parameters that influence the performance of the device is provided in Section~\ref{sec:Results}. Finally, conclusions are provided in Section~\ref{sec:conclusions}.

\section{Related Work}\label{sec:relatedwork}

One of the main problems of IoT devices is their energy consumption and their limited energy availability. When the energy of a device is depleted, it will no longer fulfill its role unless either the source of energy is replaced or a harvesting mechanism is used. Since we posit the vision of battery-less devices, energy harvesting is the only way to overcome the energy gap. It is a mechanism that allows extracting energy from external sources, such as solar irradiance, wind, thermoelectric, piezoelectric, or vibration~\cite{Shaikh2016}. 

The most used ambient-based energy harvester sources of IoT devices are radio frequency (RF), photovoltaic and thermoelectric, while the most used external sources are mechanical-based~\cite{Shaikh2016}. The decision of which type of harvester source should be used depends on the location (i.e., availability of the source) and the application requirements (e.g., consumption, execution frequency). While solar energy provides a power density of $100mW/cm^2$, RF-based harvesters are in the order of $10\mu W/cm^2$ \cite{Kim2014}. In this work, we do not consider a specific harvesting source, but model it as a generic current or voltage source.

\begin{table*}[t]\centering
\caption{Prototypes and Commercial solutions using energy harvesting}

\begin{tabular}{c c c c c}
\hline
\textbf{Platform}& \textbf{Prototype/Commercial}& \textbf{EH source}& \textbf{Radio}& \textbf{Application} \\
\hline
Takahashi \textit{et al.} \cite{Takahashi2019} & Prototype & Kinetic & 928 MHz EnOcean & Smart shoe that sends a RF tag when walking \\
Pible \cite{Fraternali2018} & Prototype & Photovoltaic & BLE  & Sense/Event occupacy \\
Dekimpe \textit{et al.} \cite{Dekimpe2019} & Prototype & RF & BLE & Smart sensor (PIR) for smart metering \\
Loubet \textit{et al.} \cite{Loubet2018}\cite{Loubet2019}  & Prototype & RF & LoRaWAN & Structural health monitoring, smart building \\
Orfei \textit{et al.} \cite{Orfei2016} & Prototype & Kinetic & LoRa & Smart metering using the bridge vibrations \\
Dalpiaz \textit{et al.} \cite{Dalpiaz2018} & Prototype & Electromagnetic & LoRa & Smart meter harvesting energy from the monitored load \\
Smart Switch$^1$ & Commercial & Kinetic & 868 MHz & Smart building, dimming/shutter control \\
Smart boot$^2$  & Commercial & Kinetic & Wi-Fi/Cellular  & Wearable and military motoring \\
Instep$^3$ & Commercial & Kinetic & BLE & Smart shoes for activity tracking \\
Bionic power Walk$^4$ & Commercial & Kinetic & - & Wearable and military motoring \\
Sequent Watch$^5$ & Commercial & Kinetic & BLE & Activity and heart rate tracking \\
Lunar Watch$^6$ & Commercial & Photovoltaic & BLE & Step counting, sleep tracking \\
Matrix Power Watch$^7$ & Commercial & Thermoelectric & BLE & Step counting, sleep tracking \\
\hline    
\label{tab:prototype}
\end{tabular}
\end{table*}

There already exist different commercial battery-less solutions available in the market, including wireless switches\footnote{\label{WirelessSwitch}“Wireless Switch”, https://www.enocean.com/en/products/product-finder/$\#$1=For+Energy+Harvesting+Wireless+Switches} that harvest kinetic energy from each push, 
smart shoes\footnote{\label{solePower}“SolePower Smartboots”, http://www.solepowertech.com/}\footnote{\label{Instep}“Instep NanoPower”, http://www.instepnanopower.com/}\footnote{\label{bionic}"Bionic PowerWalk”, https://www.bionic-power.com/} harvesting energy from foot strikes, or smartwatches powered by kinetic\footnote{\label{sequent}“Sequent Watch”, http://www.sequentwatch.com/}, solar\footnote{\label{lunar}“Lunar Watch“, https://lunar-smartwatch.com/}, or thermal\footnote{\label{matrix}“Matrix PowerWatch”, https://www.matrixindustries.com/} energy harvesting. While communications in the commercial solutions are still limited, there also exist some battery-less IoT prototypes that make use of communications (\cite{Takahashi2019}, \cite{Fraternali2018}, 
\cite{Dekimpe2019},  \cite{Loubet2018}, \cite{Loubet2019}, \cite{Orfei2016}, \cite{Dalpiaz2018}). In Table \ref{tab:prototype} we classify the listed commercial solutions and prototypes based on their energy harvesting source and we also provide their most important characteristics.

Most prior works on battery-less devices use passive RF-powered communications, such as backscatter \cite{Talla2017} \cite{Correia2016}. Instead, we focus on active communications. 
Only a few works have considered the combination of battery-less devices, energy harvesting and active radios. For example, Takahashi \textit{et al.} implemented a battery-less shoe-type wearable which generates its own electricity when walking and uses it to estimate the location of the person \cite{Takahashi2019}.
Fraternali \textit{et al.} designed a battery-less Bluetooth Low Power (BLE) sensor node that leverages ambient light and a power management algorithm to maximize the quality of service \cite{Fraternali2018}.
In \cite{Dekimpe2019}, authors present a full battery-less system that uses BLE where they stated that a capacitance above $260\mu F$ is required although many optimization tasks were needed. Since the needed capacitance will depend on the specific application and the specific requirements, in this work we will investigate those aspects to check its feasibility for LoRaWAN communications.

Loubet \textit{et al.} have been working in  LoRaWAN battery-less designs. Firstly, they presented a LoRaWAN battery-less design for a capacitor of $30mF$ \cite{Loubet2018}, but recently, they have improved the design and they were able to reduce it to $22mF$ \cite{Loubet2019}. In their measurements, every 3 seconds they can perform a cycle which includes sensing and transmitting. However, the LoRaWAN Class A standard has been optimized and does not allow any downlink, so the radio is only able to transmit data. Although in terms of power consumption this maximizes the performance, downlink is needed for many more advanced applications. For this reason, in this work we study the impact of the two downlink windows of a LoRaWAN Class A radio as well. 
The authors in \cite{Orfei2016} propose a mechanical harvester based on the vibrations of a bridge using supercapacitors. In their experiments, they test the feasibility of LoRa communications on battery-less devices. They estimated that within every transmission, less than half of the energy stored in the supercapacitor of $100mF$ is used, but even if they estimated the time required to charge the supercapacitor to its maximum voltage of $3.3V$ was 3.5 hours, it strongly depends on the number of vehicles passing along the bridge. They therefore show that supercapacitors are more designed for infrequent transmission. In this work we will provide a better insight into these values and generalized conclusions. 
Dalpiaz \textit{et al.} \cite{Dalpiaz2018} use electrical induction as a power source and a capacitor of $22mF$, to be used in smart grids. They estimate a few seconds of periodicity for measurements and transmissions, but considering energy harvesting powers up to $1465W$, which is not feasible when considering environmental energy harvesting.

Not only prototype designs have been proposed. Sherazi \textit{et al.} present a mathematical model to calculate battery life using energy harvesting in order to analyze the impact of energy harvesting sources~\cite{Sherazi2018}. Although they also use LoRaWAN in their calculations, they do assume the use of batteries.
In contrast, we consider that the harvested energy is stored in a capacitor, which has a significantly lower energy storage capacity. The smaller its capacity, the faster it will charge and discharge for a given harvester, which can lead to power failures up to several times per second~\cite{Ransford2011}.

 The battery-less vision also requires computing mechanisms to deal with this intermittent behaviour and different prototypes have already been proposed. Alpaca \cite{Maeng2017} presents a prototype that only considers static task flows, and if any task cannot be completed due to the energy level at that time, it will be executed again. AsTAR \cite{Yang2019} also presents an energy-aware task scheduler and an associated reference platform that aims to lower the burden of developing sustainable applications through self-adaptative task scheduling. It does not need any pre-configuration and supports platform heterogeneity. However, for their design they need supercapacitors up to $5F$, which might not be feasible for tiny IoT devices. Finally, Flicker \cite{Hester2017b} is an open source, open hardware prototyping platform intentionally built for battery-less applications. However, figuring out the best turn-on voltage threshold for this platform is difficult and imprecise, but an important designers aspect. For these reasons, in this work we will investigate the minimum capacitance needed and the impact on the turn on threshold that will allow to support different applications considering different environmental conditions.

Many efforts have been done to come up with solutions for the battery-less vision. However, none of these works have deeply analysed the specific requirements needed (capacitance) for specific applications (transmission interval, uplink packet size, expected downlink packet size) under specific environmental conditions (harvested power available). For this reason, we consider that this analysis is needed and that a Markov Model can help to quickly determine the requirements an applications needs.

\begin{figure*}[!t]
	\centering
	{
		\subfigure[Using a voltage source energy harvester and an ideal capacitor]{\includegraphics[width=0.67\columnwidth]{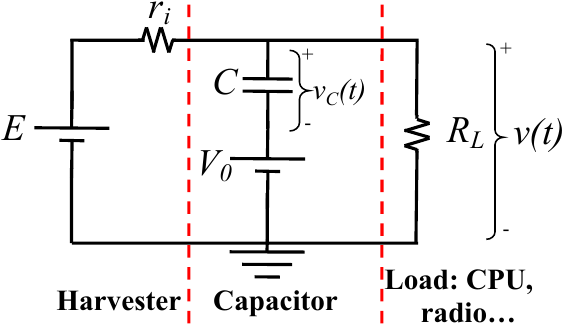}
			\label{fig:BatteryLessCircuit_SimpleVoltageHarvester}}
		\subfigure[Using a current source energy harvester and an ideal capacitor]{\includegraphics[width=0.65\columnwidth]{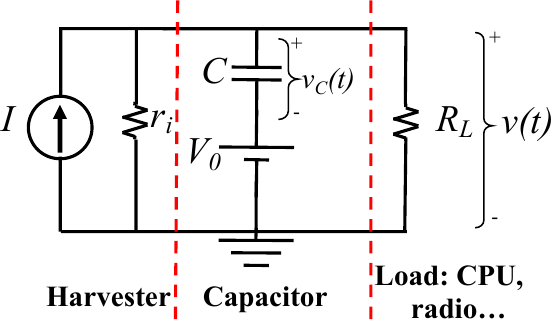}
			\label{fig:BatteryLessCircuit_SimpleCurrentHarvester}}
		\subfigure[Using a voltage source energy harvester and a real capacitor]{\includegraphics[width=0.65\columnwidth]{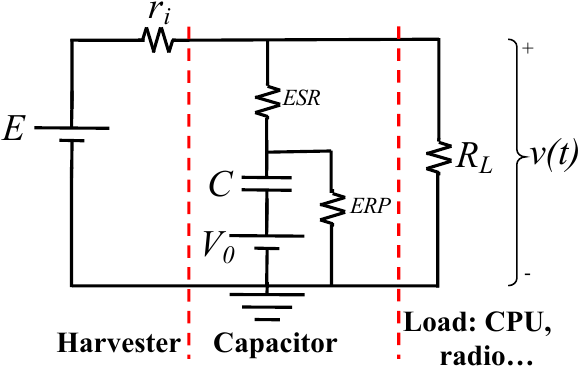}
			\label{fig:BatteryLessCircuit_CompleteVoltageHarvester}}
		
		\caption{Electrical circuit model of a battery-less IoT device}
		\label{fig:Circuits}
	}
\end{figure*}

\section{Model for a battery-less LoRaWAN device}\label{sec:model}

\subsection{Circuit Characterization}\label{sec:circuitmodel}

Battery-less IoT devices are equipped with a harvester mechanism, a capacitor, an MCU, a radio unit and the needed peripherals. In order to model the behaviour of these devices, we have considered different electrical circuits as shown in Figure~\ref{fig:Circuits}, where all of them have been divided into three main parts: the harvester (source of the energy), the capacitor (storage of the energy) and the load (consumer of the energy: MCU, radio, peripherals). More details about these 3 parts are explained below.

\paragraph{Energy Harvesting System Model} \label{sec:harvesting}

Although in reality the harvester model should depend on the type of energy source (e.g., radiation, vibration), we consider a generic and simplified approach, where only the generated power is taken into account.
Since the conditions of real harvester sources can vary over time, we have considered that the real harvester source is followed by a voltage regulator (a buck regulator as presented in \cite{sharma2018}). In this scenario, the voltage will remain constant as long as the voltage of the harvester is above a certain level. This means that, as
can be seen on the left side of Figure~\ref{fig:BatteryLessCircuit_SimpleVoltageHarvester}, the harvester is modeled as a real DC (direct current) voltage source composed of an ideal DC voltage source and a series resistance (denoted by $E$ and $r_i$, respectively).
The value of $E$ (in Volts) is chosen according to the operating voltage of the circuit elements, which in this case will be determined by the load.
The series resistance $r_i$ (in $\Omega$) limits the power of the harvester and its value is calculated using the following equation: 
\begin{equation}
 r_i = \frac{E^2}{P_{harvester}}
\end{equation} \label{eq:ri} 
where $P_{harvester}$ is the power of the harvester source, which can vary greatly depending on the type of energy harvesting considered (e.g., up to $1 mW/cm^2$ for indoor natural light, and up to $100 mW/cm^2$ for outdoor sun) \cite{Shirvanimoghaddam2018}. 

The main suitable ambient energy sources for IoT are: kinetic (motion, vibration, and mechanical), solar, thermoelectric and RF \cite{Ma2020}. Normally, thermoelectric harvesters are modeled as DC voltage sources \cite{Park2014}\cite{Carreon2014}, while RF and kinetic harvesters are modeled as an AC (alternating current) voltage source as detailed in \cite{Li2013} and \cite{Dolgov2010} for RF, and in \cite{Kim2006}, \cite{Priya2005} and \cite{Cheng2016} for kinetic sources. The AC voltage source followed by a rectifier bridge can be simplified to obtain the generic DC voltage source of Figure~\ref{fig:BatteryLessCircuit_SimpleVoltageHarvester}.
In contrast, photovoltaic cells are modeled as current harvester sources (\cite{sharma2018}, \cite{Mahmoud2012}, \cite{Saloux2011}, \cite{Dondi2007}, \cite{Dondi2008}), and for this reason, in Figure~\ref{fig:BatteryLessCircuit_SimpleCurrentHarvester} we have included its model, composed of an ideal current source and a parallel resistance (denoted by $I$ and $r_i$, respectively). Thanks to Norton's theorem of circuit theory \cite{NortonTheorem}, we can see that the presented current source energy harvester is equivalent to the voltage source energy harvester where the equivalent source current can be calculated as:

\begin{equation}
I = \frac{E}{r_i}
 \label{eq:I_Current} 
\end{equation}

\paragraph{Capacitor Model} 
The capacitor is the part of the circuit where the energy is stored. As shown in Figure~\ref{fig:behaviour}, the behaviour of the system is a succession of intervals, where the capacitor is being charged or discharged. Each interval is characterized by a specific state of the load components (e.g., MCU is active and radio is transmitting). 
We characterize the voltage of the capacitor throughout each interval using $V_0$ and $v_C(t)$. $V_0$ represents the initial voltage of the capacitor at the beginning of the interval (i.e., time $t_0$), and $v_C(t)$ is the temporal evolution of said voltage at time $t$ (relative to $t_0$). Both $V_0$ and $v_C(t)$ are included in the circuit as an ideal voltage source and the voltage over time of an ideal capacitor respectively, as shown in Figures~\ref{fig:BatteryLessCircuit_SimpleVoltageHarvester} and~\ref{fig:BatteryLessCircuit_SimpleCurrentHarvester}.

In order to consider the parasitic effects of the capacitor in the electrical circuit of a battery-less IoT device using a voltage source harvester, in Figure~\ref{fig:BatteryLessCircuit_CompleteVoltageHarvester} we have included the circuit model of a battery-less IoT device with a real capacitor, composed of an ideal capacitor with its internal resistances (both in $\Omega$): the equivalent series resistance ($ESR$) and the equivalent parallel resistance ($EPR$), which models the capacitor self-discharge.

\paragraph{Load Model} The load of the model corresponds to the set of components that consume the stored energy in the capacitor, such as 
the MCU, radio or sensors. Each of these components is characterized by a specific power consumption in each of its states (e.g., active, sleeping, off).
Therefore, they can be modeled as a load resistance denoted by $R_L$ (in $\Omega$), which can be calculated as follows:
\begin{equation}
 R_L = \frac{E}{I_{load}}
 \label{eq:RL} 
\end{equation} 

where $I_{load}$ is the sum of the supply currents of all components in their current state. $R_L$ thus varies across intervals depending on the state of each component during that interval. 

To determine if the device has enough energy at a specific time to perform its tasks (e.g., transmit data), it is needed to calculate the voltage across the load of the model $v(t)$. For the voltage source energy harvester with an ideal capacitor (cf., Figure~\ref{fig:BatteryLessCircuit_SimpleVoltageHarvester}), it can be obtained as follows:
\begin{equation}
 v(t) = E\frac{R_{eq}}{r_i}(1-e^{(\frac{-t}{R_{eq}C})})+V_0e^{(\frac{-t}{R_{eq}C})}
 \label{eq:V} 
\end{equation} 
where $C$ is the capacitance in Farads, $t$ is the time (in seconds) spent in the current interval,
and $R_{eq}$ is the equivalent resistance of the circuit (in $\Omega$), computed as:
\begin{equation}
 R_{eq}=\frac{R_L r_i}{R_L + r_i}
 \label{eq:Req} 
\end{equation} 

The value of $v(t)$ provides the voltage available in the load, which will be used to determine if a specific action (e.g., transmit, listen) can be performed during an interval, according to the needed time $t$ it will take, the energy harvesting rate $P_{harvester}$, the specific load $I_{load}$, and the capacitor voltage $V_0$ at the start of the interval. Note that $v(t)$ can be increasing or decreasing depending of the specific parameters, and if it goes below the turn-off threshold, the device (which is represented as the load in Figure~\ref{fig:BatteryLessCircuit_SimpleVoltageHarvester}) will turn off. 

The voltage available in the load $v(t)$ of the current source harvester with an ideal capacitor (cf., Figure~\ref{fig:BatteryLessCircuit_SimpleCurrentHarvester}) can be calculated as follows:

\begin{equation}
 v(t) = I R_{eq}(1-e^{(\frac{-t}{R_{eq}C})})+V_0e^{(\frac{-t}{R_{eq}C})}
 \label{eq:V_Current} 
\end{equation} 

If we combine Equation \ref{eq:V_Current} with Equation \ref{eq:I_Current}, we obtain Equation \ref{eq:V}. This means that both models, the voltage and current source energy harvester (cf., Figures \ref{fig:BatteryLessCircuit_SimpleVoltageHarvester} and \ref{fig:BatteryLessCircuit_SimpleCurrentHarvester}), are equivalent. In the rest of the paper we will focus on the voltage source model.

The introduction of the parasitic effects of the capacitor in the voltage source harvester model, modifies Equation \ref{eq:V} to:

\begin{gather}
v( t) \ =\ \frac{ER_{eq} ESR}{r_{i}( ESR\ +\ R_{eq})} e^{-\left(\frac{1}{EPR} \ \ +\ \frac{1}{ESR\ +\ R_{eq}} \ \ \right)\frac{t}{C} \ \ } + \notag \\
\frac{V_{0} R_{eq}}{ESR\ +\ R_{eq}} e^{-\left(\frac{1}{EPR} \ \ +\ \frac{1}{ESR\ +\ R_{eq}} \ \ \right)\frac{t}{C} \ \ } + \\
\frac{ER_{eq}( ESR+EPR)}{r_{i}( ESR +EPR + R_{eq})}\left( 1\ -\ e^{-\left(\frac{1}{EPR}  +\ \frac{1}{ESR\ +\ R_{eq}}  \right)\frac{t}{C}  }\right)  \notag \\
\notag
\label{eq:V_completeV} 
\end{gather}

If we define $ESR = 0$ and $EPR = \infty$, we obtain Equation~\ref{eq:V}.

\subsection{LoRAWAN Class A Device Model}\label{sec:systemmodel}

We consider a LoRa device, using the LoRaWAN medium access control (MAC) protocol. The LoRaWAN standard defines three classes of end-devices. In this work, we focus on Class A, which provides the lowest energy consumption~\cite{LoraWan} and is best suited for low power IoT devices.

\begin{figure}[t]
\centerline{\includegraphics[width=1\columnwidth]{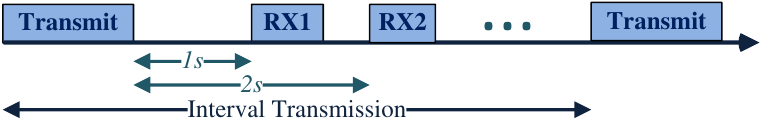}}
\caption{LoRaWAN Class A end device window timings}
\label{fig:Lora}
\end{figure}

Class A devices spend most of the time in deep sleep, and only wake up when they need to transmit data to the network server. Since they use an ALOHA-based MAC protocol, Class A LoRAWAN devices do not perform listen before talk. It is also characteristic that devices of this class are only reachable for downlink transmissions after they transmit. As can be seen in Figure~\ref{fig:Lora}, LoRaWAN Class A devices have two reception windows.
After the transmission, the device waits 1 second in the idle state and then switches to listening mode (RX1). If the device receives a downlink transmission in this first window, it does not need to stay awake for the second reception window. However, if nothing is received, it must switch again to idle mode. In that case, 2~seconds after the end of the transmission, the device will listen again (RX2) for a downlink transmission.

One of the important parameters to be configured 
is the spreading factor ($SF$), which represents the ratio between the chip rate and the baseband information rate and can range from 7 to 12. As the $SF$ increases, so does the coverage range and decoding robustness. This comes at the cost of a decrease in data rate, and thus increase in airtime. In fact, the $SF$ determines the TX time as well as the RX1 time. 
The time of RX2 is fixed, as it always uses the slowest spreading factor $SF12$.
The formulas for calculating the different times can be derived from the LoRaWAN standard~\cite{LoraWan}. First, it is convenient to define the symbol duration $T_{sym}$, in seconds, considering $SF$ bits per symbol:
\begin{equation}
 T_{sym}=\frac{2^{SF}}{BW}
 \label{eq:Tsym} 
\end{equation} 
where $BW$ is the bandwidth, which is typically 125kHz.

\begin{figure}[t]
\centerline{\includegraphics[width=0.9\columnwidth]{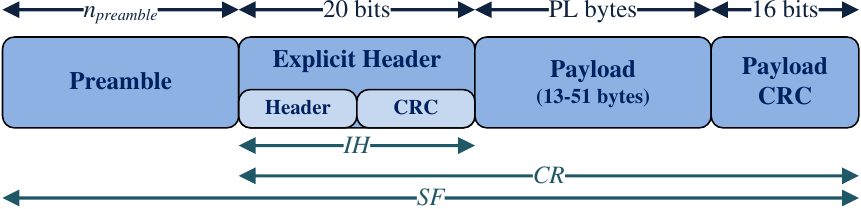}}
\caption{LoRaWAN packet format}
\label{fig:lora_packet}
\end{figure}

Figure~\ref{fig:lora_packet} shows the LoRa frame format of the physical layer. The preamble ($n_{preamble}$) is the number of programmed preamble symbols, which is 8 in LoRaWAN 1.0. Then, the low-level header can be explicitly enabled ($IH=0$) or disabled ($IH=1$). It is used to indicate the coding rate and payload length, and can be left out if both sides of the communication have these parameters fixed. The time needed to transmit or receive the preamble sequence is given by:
\begin{equation}
 T_{preamble}=(n_{preamble} + 4.25) \cdot T_{sym}
 \label{eq:Tpreamb} 
\end{equation} 

\begin{figure}[t]
\centerline{\includegraphics[width=0.9\columnwidth]{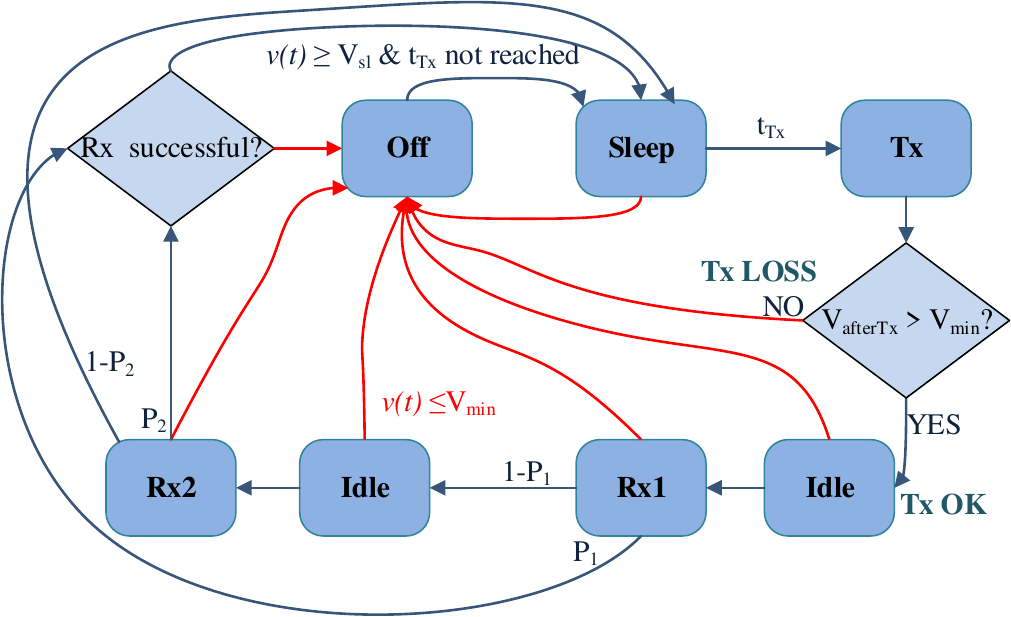}}
\caption{State diagram of a battery-less LoRaWAN Class A device}
\label{fig:blocks}
\end{figure}

The number of symbols that make up the packet payload and header are given by Equation~\ref{eq:PLSym}, and the payload duration is given by Equation~\ref{eq:Tpayload}.
\begin{multline}
S_{payload} = 8 + \\
 max (\ceil[\Bigg]{ \frac{8PL - 4SF + 28 +16 - 20 IH}{4(SF - 2DE)}} (CR + 4), 0)
 \label{eq:PLSym} 
\end{multline} 
\begin{equation}
 T_{payload} = S_{payload} \cdot T_{sym}
 \label{eq:Tpayload} 
\end{equation}
where $PL$ is the number of payload bytes (which can vary from 13 to 51), $CR$ is the coding rate, which refers to the proportion of transmitted bits that actually carry information (typical values are 4/5, 6/8 or 4/8, and higher values mean more overhead), and the low data rate optimization can be enabled with $DE =1$ and disabled with $DE=0$ (which intends to correct the clock drift at $SF11$ and $SF12$). 
Finally, the time on air (or packet duration) can be calculated as follows:

\begin{equation}
 T_{p} = T_{preamble} + T_{payload}
 \label{eq:Tpacket} 
\end{equation}

\subsection{Device State Transitions}\label{sec:Simulation Model}
The main problem of battery-less devices is how to deal with their intermittent behaviour, and the energy and time needed to be awake to perform the different actions (e.g., listen, transmit, receive). 
As shown in Figure~\ref{fig:behaviour}, an intermittent device has two main states: On and Off. In this work we assume that in these two states, the device is constantly harvesting energy. Specifically, when the device is in the On state, it can be in one of the following sub-states: Sleep, Idle, Tx, Listen, or Rx. In these cases, it will stay turned on until it actively turns itself off, or when the capacitor voltage drops below the turn-off threshold $V_{min}$. This is the lowest voltage at which its radio and MCU can safely operate. When the device is in the Off state, it will stay turned off until its capacitor reaches the turn-on voltage threshold $V_{sl}$, which is a configurable parameter of the model. 
Depending on the $P_{harvester}$ and the power consumption of the MCU (which determines $R_L$), the capacitor can be charging or discharging during sleep mode (cf., Equation~\ref{eq:V}).

The complete state diagram of the simulation system is shown in Figure~\ref{fig:blocks}. 
Whenever the device has something to transmit, it will be checked in which state the system is. If the system is in Off mode, it will not be able to transmit and the packet opportunity will be lost. However, if the system is in the Sleep state, it will try to send it. If the current voltage in the load is enough to transmit it, it will do so and will switch to the Idle state for 1 second (as shown in Figure~\ref{fig:Lora}). In order to know the needed voltage to transmit, Equation~\ref{eq:V} is used, where the time $t$ corresponds to the time on air (or time needed to send the packet), which can be calculated with Equation~\ref{eq:Tpacket}. After being 1 second in the Idle state, the system will switch to Listen, in the first reception window (RX1), where the device checks whether a preamble is received. If a preamble is detected, which will happen with probability $P_1$, the device continues receiving the downstream transmission. The time spent during this reception is calculated with Equation~\ref{eq:Tpacket} and the corresponding $SF$. After receiving the packet, the device will switch to Sleep mode. 

However, if a preamble has not been detected in the first window (with probability $1 - P_1$) during the time determined by Equation~\ref{eq:Tpreamb} (where the $SF$ is the corresponding spreading factor used in the transmission), it closes the receive window and transitions to the second Idle state. 
2 seconds after the end of the transmission, the system will switch to the second reception window (RX2), where the device will check again if a preamble is received. In this case, with probability $P_2$, a packet is received. The time needed to receive the packet is calculated with Equation~\ref{eq:Tpacket}  but using an $SF$ value of 12. However, if nothing is received (with probability $1 - P_2$) during the time determined by Equation~\ref{eq:Tpreamb} (using an $SF$ value of 12), the device will just switch to Sleep.
At any point, the system will switch to the Off state immediately whenever the voltage of the capacitor reaches the $V_{min}$ value, and any ongoing transmission or reception will fail.

\section{Markov Chain Model}\label{sec:MC model}
In order to evaluate the system performance for different values of the parameters, 
we propose a Markov Chain Model of the described battery-less LoRaWAN device. 
This model takes as inputs the current consumption ($I_{load}^{ST}$) of the different device states, the capacitance ($C$), the energy harvesting power ($P_{harvester}$), how often a transmission will take place (interval transmission time, $M$), the probability of receiving a downlink packet in the first and in the second window ($P_1$ and $P_2$ respectively), the uplink and downlink packet size ($PS$ and $DLPS$), and the Spreading Factor ($SF$) to be used. 
As for the outputs, the Markov Chain Model determines the uplink Packet Delivery Ratio ($PDR$) achievable and the probabilities of successfully receiving downlink packets in the first or in the second window ($PDL_1$ and $PDL_2$ respectively).
We make two assumptions to be able to model the problem like a discrete process: 
time is divided into equal length time slots of length 1 time unit, and also, the energy level of the device is assumed to have a discrete and finite range of values. 

\subsection{Device and transmission related parameters and assumptions}\label{sec:MC assumptions}

We consider a battery-less LoRaWAN device that can be in six different states: Off, Sleep, Idle, Tx, Listen and Rx, 
which correspond with the states discussed in Section \ref{sec:Simulation Model}. In order to simplify the model, we consider the voltage source model with an ideal capacitor (Figure \ref{fig:BatteryLessCircuit_SimpleVoltageHarvester}) in this section. The energy function for every state is given by the discrete version of Equation \ref{eq:V}:

\begin{equation}
 \hat V(ST, \hat V_0, t) =  E\frac{R_{eq}^{ST}}{r_i} (1-e^{(\frac{-t}{R_{eq}^{ST}C})})+ \hat V_0 e^{(\frac{-t}{R_{eq}^{ST}C})}
 \label{eq:V_STATE} 
\end{equation}
where $ST$ represents the specific state, $\hat V_0$ is the discrete voltage at 
a certain point in time while being in state $ST$ and $t$ is the time elapsed since that certain point in time.
The values for $R_{eq}^{ST}$ are specifically chosen according to the state $ST$, i.e. Equations \ref{eq:RL} and \ref{eq:Req} have been used to get this value and $I_{load}^{ST}$ is the current consumption of the specific state $ST$. 
We assume that the parameter values of $I_{load}^{ST}$ are such that the voltage functions $\hat V(ST, \hat V_0, t)$ (i.e., Equation \ref{eq:V_STATE}) are increasing for the Off, Sleep and Idle states and decreasing for Tx, Listen and Rx.

Finally, the time (in seconds) needed to get a voltage level $\hat V_f$, starting from $\hat V_i$, for a specific state $ST$ can be computed using Equation \ref{eq:V_STATE} as:

\begin{equation}
 t (ST, \hat V_i, \hat V_f) = -R_{eq}^{ST}C\ln(\frac{\hat V_f - E\frac{R_{eq}^{ST}}{r_i}}{\hat V_i - E\frac{R_{eq}^{ST}}{r_i}})
 \label{eq:T_STATE} 
\end{equation} 

Table~\ref{tab:MCmodelparameters} summarizes the Markov Chain parameters used. It is important to note that when at instant $n$, the state is Sleep and the energy level is $\hat V(n)$ but such that $\hat V(Tx, \hat V(n), T_{Tx}) \leq \hat V_{min}$, then the transmission will take place but will be aborted when the energy level reaches $\hat V_{min}$ and a switch to Off will be made. Hence in this case the transmission is not successful and energy is wasted. After the aborted transmission, the system is in the Off state and the energy level is $\hat V_{min}$. The same assumption holds for receiving a packet: if $\hat V(Rx, \hat V(n), T_{Rx}) \leq \hat V_{min}$, then the reception of the packet will take place but will be aborted when the energy level reaches $\hat V_{min}$ and a switch to Off will be made.

\begin{table}[t]
\begin{footnotesize}
\caption{Parameters of the Markov Chain Model}
\begin{center}
\begin{tabular}{r l}
\hline
\textbf{Symbol}& \textbf{Meaning}\\
\hline
$n$ & Markov Chain time instant\\
$M$ & Number of time units between two consecutive transmissions\\
$\hat V_{sl}$ & Voltage level where the device makes the switch Off-Sleep,\\
	& the turn-on voltage threshold \\
$\hat V_{min}$ & Voltage level under which the system is in the OFF state \\
$T_{Tx}$ &  Time needed to transmit a packet \\
$T_{Rx1}$ & Time needed to receive a packet in the first window \\
$T_{Rx2}$ & Time needed to receive a packet in the second window \\
$T_{L1}$ & Time needed for listening to the first preamble\\
$T_{L2}$ & Time needed for listening to the second preamble\\ 
$T_{Id1}$ & First Idle waiting period (1sec) \\ 
$T_{Id2}$ & Second Idle waiting period  \\
$P_1$ & Probability of receiving a downlink packet in the first \\ 
& window \\
$P_2$ & Probability of receiving a downlink packet in the second \\ 
& window \\
$\hat V_{Tx}$ & Minimal voltage level when in Sleep  at which a \\ 
& transmission is still successful: $\hat V(Tx, \hat V_{Tx},T_{Tx}= \hat V_{min}+1$\\
$\hat V_{Rx1}$ & Minimal voltage level that is necessary to receive a packet \\
& successfully in the first window: \\ &$\hat V(Rx, \hat V_{Rx1}, T_{Rx1})= \hat V_{min}+1$\\
$\hat V_{Rx2}$& Minimal voltage level that is necessary to receive a packet \\ 
& successfully in the second window: \\ &$\hat V(Rx, \hat V_{Rx2}, T_{Rx2})= \hat V_{min}+1$\\
\hline
\end{tabular}
\label{tab:MCmodelparameters}
\end{center}
\end{footnotesize}
\end{table}

As explained in Section \ref{sec:model}, $T_{Tx}$, which is the time needed to transmit a packet can be calculated with Equation~\ref{eq:Tpacket} with the corresponding $SF$. $T_{Id1}$ equals to 1 second and represents the time the radio spends in the Idle state between the end of the transmission and the first reception window. $T_{L1}$ is the time needed for listening to the first preamble and can be calculated using Equation~\ref{eq:Tpreamb}, where the $SF$ is the corresponding spreading factor used in the transmission. Then, the second idle period takes place, $T_{Id2}$, and can be calculated as $1-T_{L1}$. The time needed for the second listening window is represented by $T_{L2}$ and is determined by Equation~\ref{eq:Tpreamb} but using an $SF$ value of 12. In case a reception is detected, $T_{Rx1}$ and $T_{Rx2}$ represent the time needed to receive a packet in the first and second window respectively. Their values can be calculated using Equation~\ref{eq:Tpacket}, where in the first case the corresponding $SF$ of the transmission is used, and in the later case, $SF=12$.

In order to be able to guarantee the successful states, i.e. transmit or receive a packet, the minimal needed voltage levels are defined as $\hat V_{Tx}$, $\hat V_{Rx1}$ and $\hat V_{Rx2}$. $\hat V_{Tx}$ corresponds to the minimal voltage level
at which a transmission is still successful, which can be represented as $\hat V(Tx, \hat V_{Tx},T_{Tx})= \hat V_{min}+1$. $\hat V_{Rx1}$ is the minimal voltage level that is necessary to receive a packet successfully in the first window, and can be represented as $\hat V(Rx, \hat V_{Rx1}, T_{Rx1})= \hat V_{min}+1$. Finally, $\hat V_{Rx2}$ is the minimal voltage level that is necessary to receive a packet successfully in the second window, and can be represented as $\hat V(Rx, \hat V_{Rx2}, T_{Rx2})= \hat V_{min}+1$.


We consider a sensor application where the device needs to send a measurement periodically. For this reason, we let the number of time slots between two consecutive transmission instants be constant and equal to $M$ time units. For modelling reasons, we let this interval be larger than the time to transmit Uplink (UL) packets and to receive Downlink (DL) packets, i.e.:

\begin{equation}
M > T_{Tx}+ T_{Id1} + T_{L1} + T_{Id2} + max(T_{Rd2}, T_{L2}) 
 \label{eq:M_requirements} 
\end{equation} 

\begin{figure}[t]
\centerline{\includegraphics[width=0.8\columnwidth]{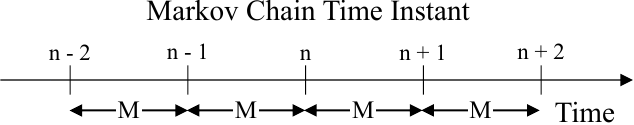}}
\caption{Markov Chain time instants}
\label{fig:MCtimeinstants}
\end{figure}

\subsection{Markov System Model} \label{sec:MC System Model}

We consider the system at the scheduled transmission instants, denoted by $..., n-1, n, n+1,...$, 
and calculate if a successful transmission occurs or not. Between each $n$ and $n+1$, there are $M$ time steps. 
 Figure~\ref{fig:MCtimeinstants} shows the transmission instants of the Markov Chain. The state of the system at time instant $n$ is described by 
$\hat V(n)$, the voltage level
, and $S(n)$, the state of the system
, where $S(n) \in \{OFF, SL0, SL1\}$. 
Note that the system states are denoted in capital letters (OFF, SL0, SL1) while the device states defined in Section \ref{sec:Simulation Model} are denoted as Off, Sleep, Idle, Tx, Listen, and Rx.

\begin{figure*}[t]
\centerline{\includegraphics[width=0.8\linewidth]{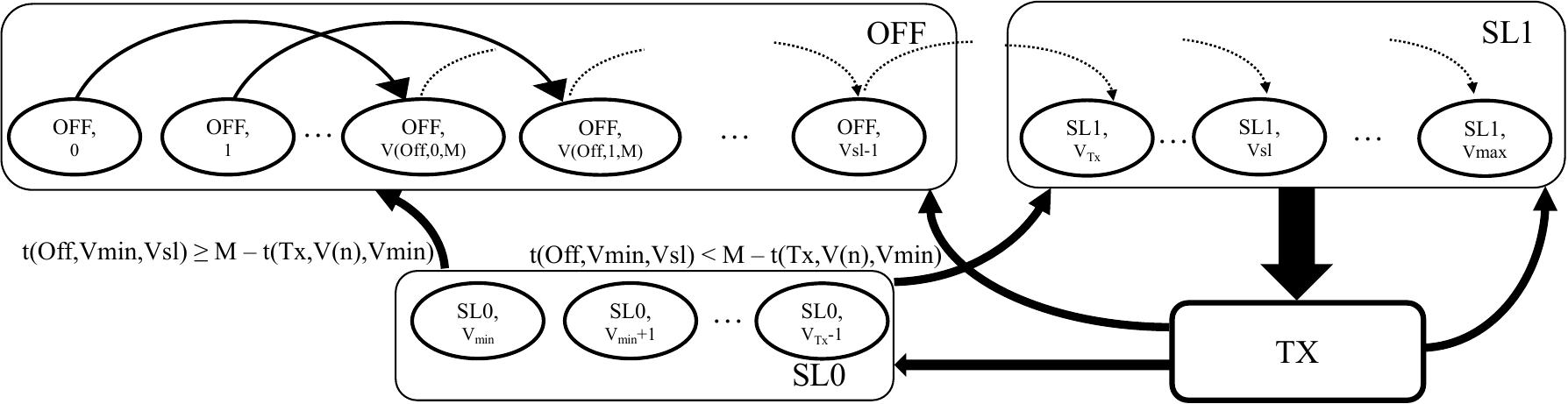}}
\caption{Markov Chain Diagram, where transitions between the three main states ($OFF$, $SL0$ and $SL1$) are shown. The specific transitions of $SL1$ are represented in Figure~\ref{fig:MC_SL1}}
\label{fig:MCDiagram}
\end{figure*}

The process $(\hat V(n), S(n))$ is a discrete-time discrete-value Markov Chain, and can be categorized into three main states, as can be seen in Figure~\ref{fig:MCDiagram}, where the Markov Chain Diagram is shown:
\begin{itemize}
\item $OFF$: this state corresponds with the blue areas of Figure~\ref{fig:behaviour} and $0 \leq \hat V(n) < \hat V_{sl}$.
\item $SL0$: this state corresponds with the pink areas of Figure~\ref{fig:behaviour} and $\hat V_{min} \leq \hat V(n) < \hat V_{Tx} $, which means that there is not enough energy to perform a transmission.
\item $SL1$: this state corresponds with the pink areas of Figure~\ref{fig:behaviour} and $\hat V_{Tx} \leq \hat V(n) < \hat V_{max}$, which means that there is enough energy to perform a transmission.
\end{itemize}

In order to compute the steady state of this Markov Chain, we need to compute all possible transition probabilities:

\paragraph{\textbf{If $\textbf{S(n) = OFF}$}} The transmission of the packet that is scheduled at time instant $n$ is not transmitted and hence considered lost. 
If $t(Off, \hat V(n), \hat V_{sl}) \geq M$, the system will stay in the OFF state at $S(n+1)$ and $\hat V(n+1) =  \hat V(Off, \hat V(n), M)$. Otherwise, $S(n+1) = SL1$ and $\hat V(n+1) =  \hat V(Sleep, \hat V_{sl}, M - t(Off, \hat V(n), \hat V_{sl}))$.

\paragraph{\textbf{If $\textbf{S(n) = SL0}$}} The transmission of the packet that is scheduled at time instant $n$ is started, but due to lack of energy, it is aborted and hence, considered lost. The state at $n+1$ will depend on whether the voltage level at $n+1$ reaches the level $\hat V_{sl}$  or not. 
If $t(Off, \hat V_{min} , \hat V_{sl}) \geq M - t(Tx, \hat V(n) , \hat V_{min})$, then the system will be in the OFF state at $S(n+1)$ and $\hat V(n+1) =  \hat V(Off, \hat V_{min}, M - t(Tx, \hat V(n) , \hat V_{min}))$. Otherwise, it will switch to the SL1 state and $\hat V(n+1) =  \hat V(Sleep, \hat V_{sl}, M - t(Tx, \hat V(n) , \hat V_{min}) -t(Off, \hat V_{min} , \hat V_{sl}) )$.

\begin{figure}[t]
\centerline{\includegraphics[width=1\linewidth]{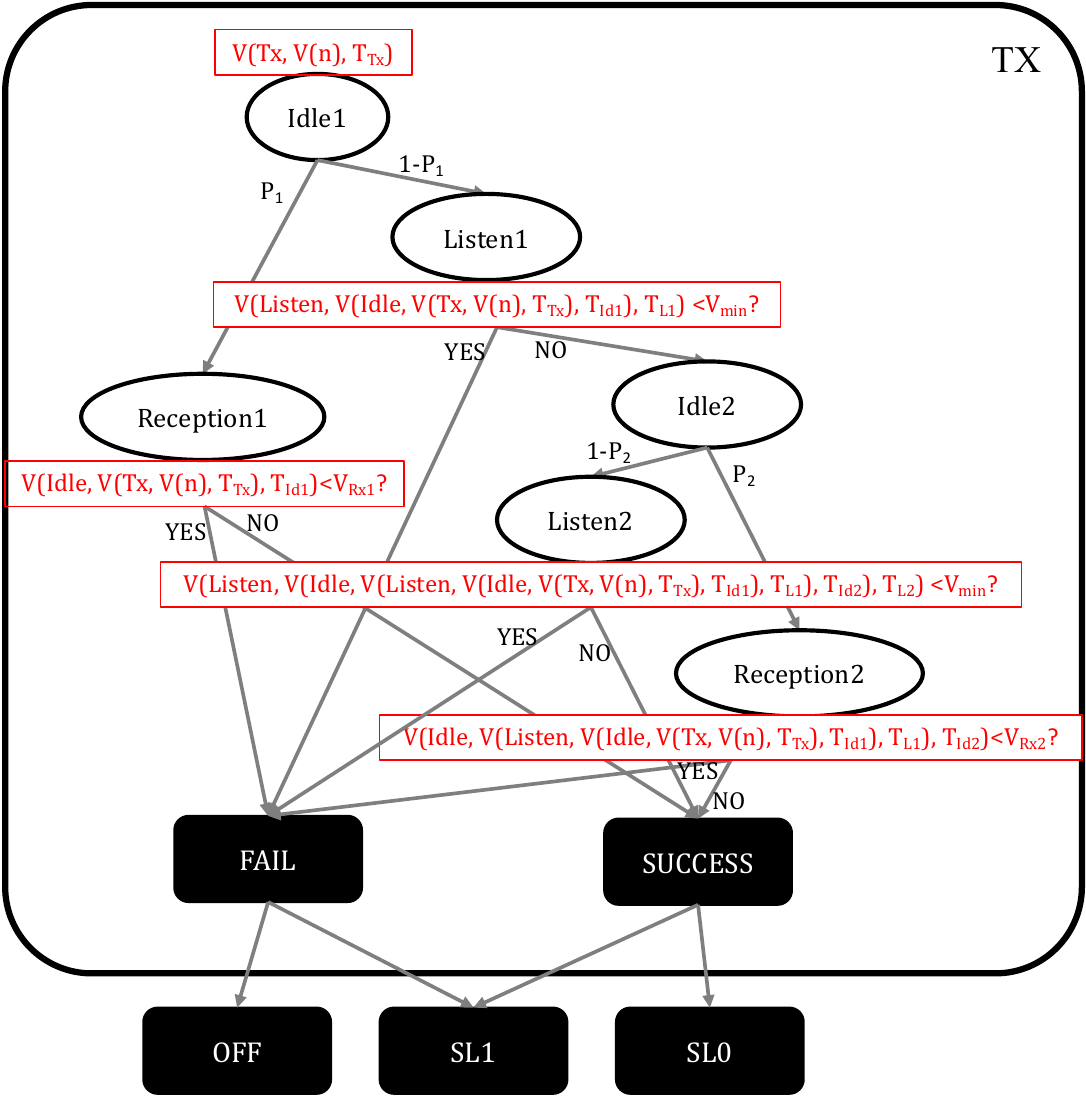}}
\caption{Transitions within SL1, which are those where the transmission of the packet has been possible and depending on the specific Voltage and the probabilities of reception, the next state will be determined}
\label{fig:MC_SL1}
\end{figure}

\paragraph{\textbf{If $\textbf{S(n) = SL1}$}} The transmission that is scheduled at time instant $n$ is transmitted successfully since $\hat V_{Tx} \leq \hat V(n)$. Figure~\ref{fig:MC_SL1} shows all the transition probabilities within this $SL1$ state. The first Idle period of 1 second starts immediately after the transmission. After $T_{Id1}$, the voltage level is given by $\hat V(Idle, \hat V(Tx, \hat V(n), T_{Tx}), T_{Id1})$. We consider two possibilities: the reception of a DL packet is started (with probability $P_1$) or not (with probability $1 - P1$). 

For example, as shown in Figure~\ref{fig:MC_SL1}, if the reception of a DL packet is started (probability $P_1$), it can be successful or not, depending on the available voltage at the start of the reception period. 
If $\hat V(Idle, \hat V(Tx, \hat V(n), T_{Tx}), T_{Ix1}) \geq \hat V_{Rx1}$, then the DL packet is received successfully and after the reception, the voltage level is given by $\hat V(Rx, \hat V(Idle, \hat V(Tx, \hat V(n), T_{Tx}), T_{Id1}), \hat V_{Rx1})$, which is the voltage after performing the cycle of transmitting a packet, wait 1 second in the Idle state and then receiving a packet in the first reception window. 
The system state at time $n+1$ can be $SL0$ if $\hat V(Sleep, \hat V(Rx, \hat V(Idle, \hat V(Tx, \hat V(n), T_{Tx}), T_{Id1}), \hat V_{Rx1}),$ $ M-T_{Tx}-T_{Ix1}- T_{Rx1}) < \hat V_{Tx}$ (i.e. the voltage at the time the next transmission needs to be done is smaller than the needed voltage to perform such transmission), or $SL1$ otherwise.
However, if $\hat V(Idle, \hat V(Tx, \hat V(n), T_{Tx}), T_{Id1}) < \hat V_{Rx1}$, then the DL packet reception is aborted due to lack of energy, and the next state could be $OFF$ if there is not enough time before the next event to achieve the turn-on threshold $\hat V(t(Off, \hat V_{min} , \hat V_{sl}) \geq M - T_{Tx} - T_{Id1} - t(Rx, \hat V(Idle, \hat V(Tx, \hat V(n), T_{Tx}), T_{Id1}), \hat V_{min})$), or $SL1$ otherwise.
For the rest of the cases, the same strategy is followed, as shown in Figure~\ref{fig:MC_SL1}.

\subsection{Packet Delivery Ratio Calculation}\label{sec:PDR}
\begin{figure}[t]
\centerline{\includegraphics[width=1\linewidth]{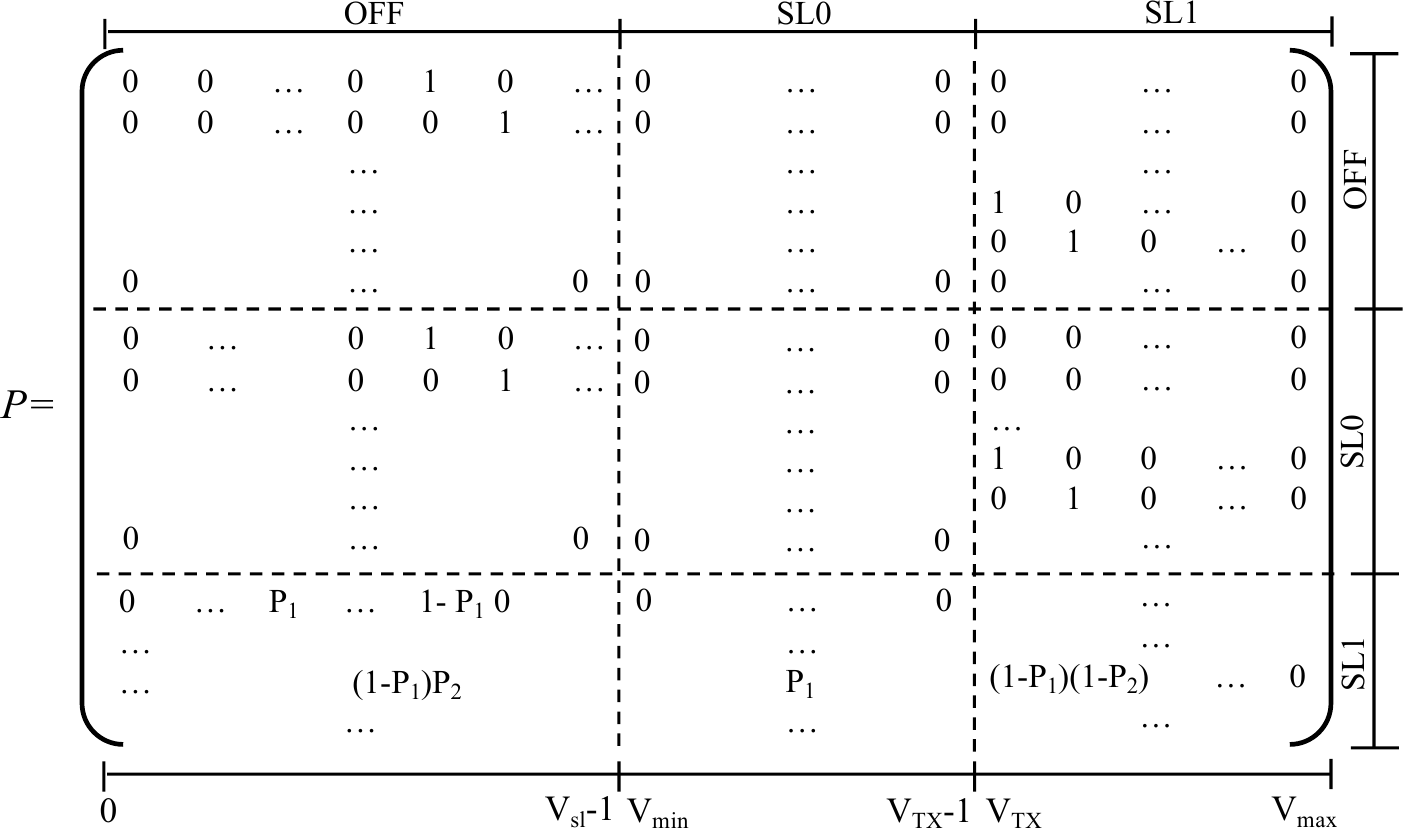}}
\caption{Example of a transition matrix $P$, where each row and column represent the set of possible states of the process $(\hat V(n), S(n))$}
\label{fig:MC_P}
\end{figure}

The process $(\hat V(n), S(n))$ is fully characterized by its transition matrix $P$, where each element in the matrix $p^{i,j}$ refers to the probability by which the transition from state $i$ to state $j$ occurs. The size of the matrix $P$ defines the size of the space state, and therefore the complexity of the computations. Figure~\ref{fig:MC_P} shows an example of the transition matrix $P$, where the row is the source and the column is the destination state (which are characterized by the system state and voltage). 
Most states transition to one other state with probability 1, as could be seen in Figure~\ref{fig:MCDiagram} where only one arrow goes from one state to another one. However, and as it is shown in Figure~\ref{fig:MC_SL1}, within $SL1$, the transition probabilities are dependant on $P_1$ and $P_2$, and this is represented in the bottom part of the example transition matrix $P$ of Figure~\ref{fig:MC_P} .

 Once the transition matrix $P$ is determined based on the voltages and probabilities, we can compute the steady state vector $\pi$ at the scheduled transmission time instants with components $\pi(k,l)$, where $k$ takes the values $OFF$, $SL0$, or $SL1$, and $l$ is the voltage level in these states. Since $\pi$ is the steady vector, it satisfies:

\begin{equation}
\pi \cdot P = \pi \quad\text{and}\quad \pi \cdot e = 1
 \label{eq:MC_satis1} 
\end{equation} 
where $e$ is the unit vector. 
Once $\pi(k,l)$ is determined, it is possible to compute the Packet Delivery Ratio (PDR) of UL packets:

\begin{equation}
PDR = 1 - (\sum_{l=1}^{\hat V_{sl} - 1} \pi(OFF,l) + \sum_{l=1}^{\hat V_{Tx} - 1} \pi(SL0,l) )
 \label{eq:MC_PDR} 
\end{equation}

It is also possible to determine the probability of successfully receiving DL packets in the first reception window:

\begin{equation}
PDL_1= \sum_{l=\hat V_{Tx}}^{\hat V_{max} - 1} P_1 \cdot \pi(SL1,l)\cdot  \delta(\hat V1_l \geq \hat V_{Rx1})
 \label{eq:MC_lossDL1} 
\end{equation} 
where $\delta(x) = 1 $ if the condition $x$ is satisfied and 0 otherwise, and $\hat V1_l = \hat V(Idle, \hat V(Tx, l,T_{Tx}), T_{Id1})$. And the probability of successfully receiving DL packets in the second reception window is given by:

\begin{equation}
PDL_2= \sum_{l=\hat V_{Tx}}^{\hat V_{max} - 1} (1-P_1) P_2 \cdot \pi(SL1,l)\cdot  \delta(\hat V2_l\geq \hat V_{min})
 \label{eq:MC_lossDL2} 
\end{equation} 
where $\hat V2_l = \hat V(Idle, \hat V(Listen, \hat V(Idle,\hat  V(Tx, l, T_{Tx}),$ $T_{Id1}), T_{L1}), T_{Id2})$, and corresponds to the voltage level the device has before the second window reception, when the transmission has started at the voltage level $l$.

\section{Evaluation of the Battery-less IoT device model} \label{sec:modelevaluation}

\subsection{Simulation Setup}\label{sec:simulationsetup}
In this section, we evaluate the battery-less IoT device models presented in Section \ref{sec:model}. The simulation environment has been implemented as an event-based simulator in C++. We have implemented the LoRaWAN Class A MAC protocol, that assumes the LoRa physical layer, where parameters such as the bandwidth, coding rate or number of preamble symbols can be defined. On the other side, it also includes the energy module to model the capacitor voltage. The energy consumption of the states are based on the Semtech SX1272/73 LoRa radio \cite{Semtech}. 
As such, $V_{min}$ and $E$ have been defined as 1.8V (minimum operating voltage of the SX1272/73) and 3.3V (typical operating voltage), respectively.
For all the cases, we assume periodic uplink transmissions at a constant interval.
The transmission power of the radio has been set to +13 dBm, as stated in Table~\ref{tab:simStates}. For simplicity of the analysis, we assume a constant energy harvesting rate during a single experiment, which is in line with the output of a buck regulator, as explained in Section \ref{sec:model}. And in this analysis, we do not consider the energy consumed by sensors (which in many cases is negligible compared to energy consumed by the radio), or collisions between devices is left for future work. Table~\ref{tab:generalparameters} summarizes the general parameters used in the simulations.

\begin{table}[t]
\begin{footnotesize}
\caption{States of the system (derived from \cite{Semtech} and \cite{microcontroller})}
\begin{center}
\begin{tabular}{l c c r}
\hline
\textbf{System State}& \textbf{MCU State}& \textbf{Radio State} & $R_{L}$ \\
\hline
Off & Off &  Off  &  600 $k\Omega$ \\
Sleep & Sleep & Sleep  &  589.286 $k\Omega$\\
Idle & Sleep & Idle  &  471.428 $k\Omega$\\
Tx (+13 dBm) & Active &  Tx & 117.811 $\Omega$ \\
Listen & Active & Listen &  313.957 $\Omega$ \\
Rx & Active & Rx  &  294.354 $\Omega$\\

\hline
\end{tabular}
\label{tab:simStates}
\end{center}
\end{footnotesize}
\end{table}

\begin{table}[t]
\begin{footnotesize}
\caption{General simulation parameters}
\begin{center}
\begin{tabular}{l l r}
\hline
\textbf{Parameter}& \textbf{Symbol}& \textbf{Value} \\
\hline
min Voltage & $V_{min}$ & 1.8V \\
Operating Voltage & $E$ & 3.3V \\
Coding Rate & $CR$ & 4/5 \\
Bandwidth & $BW$ & 125kHz \\
Preamble symbols & $n_{preamble}$ & 8\\
Data rate optimization enabled & $DE$ & 0\\
Header disabled & $IH$ & 1\\
\hline
\end{tabular}
\label{tab:generalparameters}
\end{center}
\end{footnotesize}
\end{table}

As explained before, the values used for the load depend on the specific state of the load components. We consider that the load is formed by the radio and the MCU (i.e., an STM32L162xE chip \cite{microcontroller}). 
In this case, current consumption in  Low-power run mode (Active) and in Low-power sleep mode (Sleep) are considered. Table~\ref{tab:simStates} defines the considerations of the MCU and the radio for the different states of the system and the corresponding values for the load, $R_L$, which are determined using Equation~\ref{eq:RL}.

\subsection{Model parameters evaluation}	\label{sec:parameterseva}
In Section \ref{sec:model}, we introduced three different battery-less IoT device models: (i) a voltage source energy harvester with an ideal capacitor, (ii) a current source energy harvester with an ideal capacitor and (iii) a voltage source energy harvester with a real capacitor. While we already demonstrated that the voltage source energy harvester is equivalent to the current source energy harvester model above, we want to evaluate the effect of the parasitic resistances with the capacitor in this section.

The values of $ESR$ and $EPR$ determine the behaviour of the real capacitor. $ESR$ is usually defined in the data sheet of the capacitors. However, $EPR$ models the capacitor self-discharge. Capacitor data sheets usually give this value as the leakage current. Using Ohm’s law we can easily use this to determine the value of $EPR$. These two values are very technology dependent, and normally, the bigger the capacitance is, the worse they are. In order to evaluate the effects of $ESR$ and $EPR$ on performance, we have considered three off-the-shelf supercapacitors of 1F (the maximum capacitance considered in this paper).

First, we have considered an 
SCCQ12E105PRB\footnote{https://www.mouser.es/datasheet/2/40/AVX-SCC-3.0V-1128335.pdf} capacitor, with a $ESR = 1.5 \Omega$ and a $EPR = 550000 \Omega$. Cap-XX supercapacitors are the smallest devices available for a given ESR and capacitance (high power and energy density). For this reason, we have considered two of their capacitors. The 
DMT3N4R2U224M3DTA0\footnote{https://www.tecategroup.com/products/datasheets/cap-xx/CAP-XX\%20DMT220mF\%20Datasheet\%20Rev\%201-2.pdf} capacitor, with a $ESR = 0.36 \Omega$ and a $EPR = 1100000 \Omega$, and 
DMF4B5R5G105M3DTA0\footnote{https://www.tecategroup.com/products/datasheets/cap-xx/CAP-XX\%20DMF1F\%20Datasheet\%20Rev\%201-2.pdf}, with a $ESR = 0.05 \Omega$ and a $EPR = 550000 \Omega$.

\begin{figure}[t!]
\centerline{\includegraphics[width=1\columnwidth]{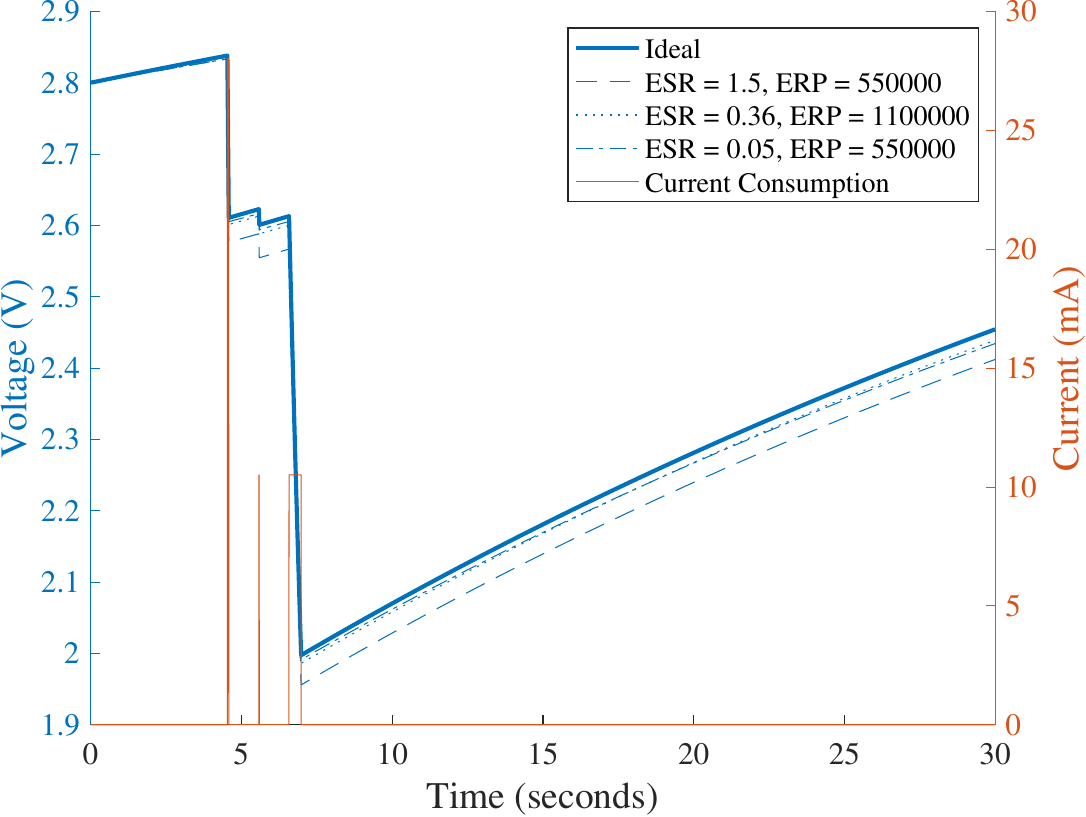}}
\caption{Transmission behaviour for a $C=4.7mF$ when no DL is received for an UL PS of 16B using $SF$7 and $P_{harvester}=1mW$ }
\label{fig:tx1cycle}
\end{figure}

Figure \ref{fig:tx1cycle} shows the voltage and current consumption variation over time for a transmission cycle of a LoRaWAN class A end device. The uplink packet size is set to 16 bytes using Spreading Factor 7 and no downlink is expected. We have considered a $P_{harvester}=1mW$, and an ideal and three real capacitor of $4.7mF$. The three real capacitors have the parasitic resistances described above. As can be seen the transmission starts at 4.53 seconds, and has a current consumption of 28 mA, which implies a considerable voltage drop. During the two idle states of the transmission, the voltage increases. During the two reception windows the voltage drops again. This drop is more significant in the second reception window because it is longer, lasting around 400ms and consuming more than 10 mA.

We can also see how the voltage variation over time for the real capacitor depends on its parasitic resistances $ESR$ and $EPR$. The lower the value of $ESR$ and the higher the value of $EPR$ are, the better are the obtained results. In fact, if we consider $ESR = 0$ and $EPR = \infty$, we obtain the Ideal results. For the rest of the paper we consider the ideal capacitor, since the self-discharge effects are very technology dependant and can be easily accounted for in our model. Moreover, as capacitors considered for IoT devices are relatively small, the effects of self-discharge on the results are minimal.

\section{Accuracy of the Markov Chain model} \label{sec:accuracy}

	\begin{figure*}[t]
	\centering

	{
		\subfigure[Simulator vs. MC Model with granularity 100]{\includegraphics[width=0.65\columnwidth]{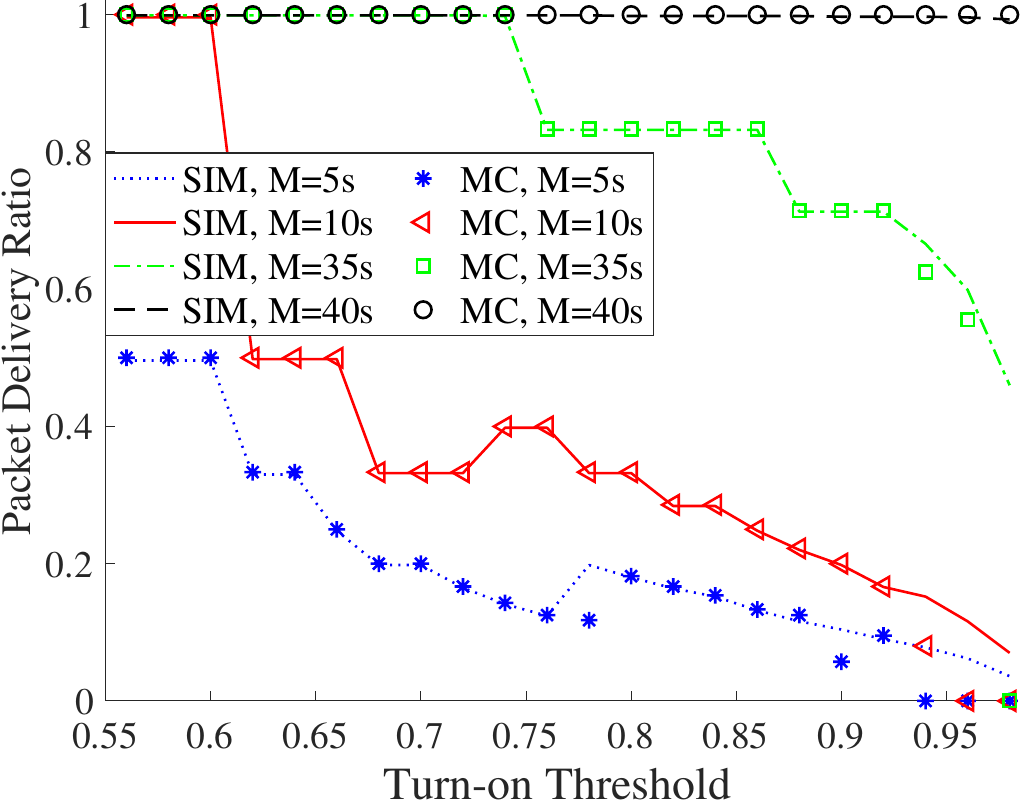}
			\label{fig:1dA_0_0_gran100}}
		\subfigure[Simulator vs. MC Model with granularity 500]{\includegraphics[width=0.65\columnwidth]{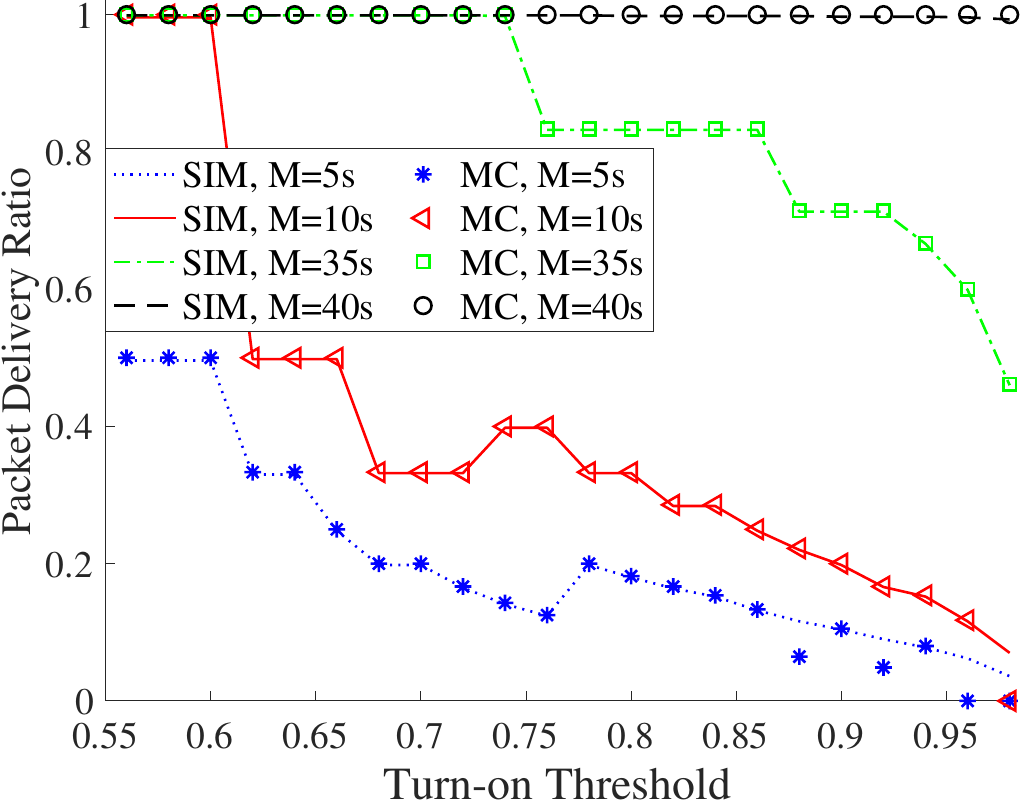}
			\label{fig:1dA_0_0_gran500}}
		\subfigure[Simulator vs. MC Model with granularity 750]{\includegraphics[width=0.65\columnwidth]{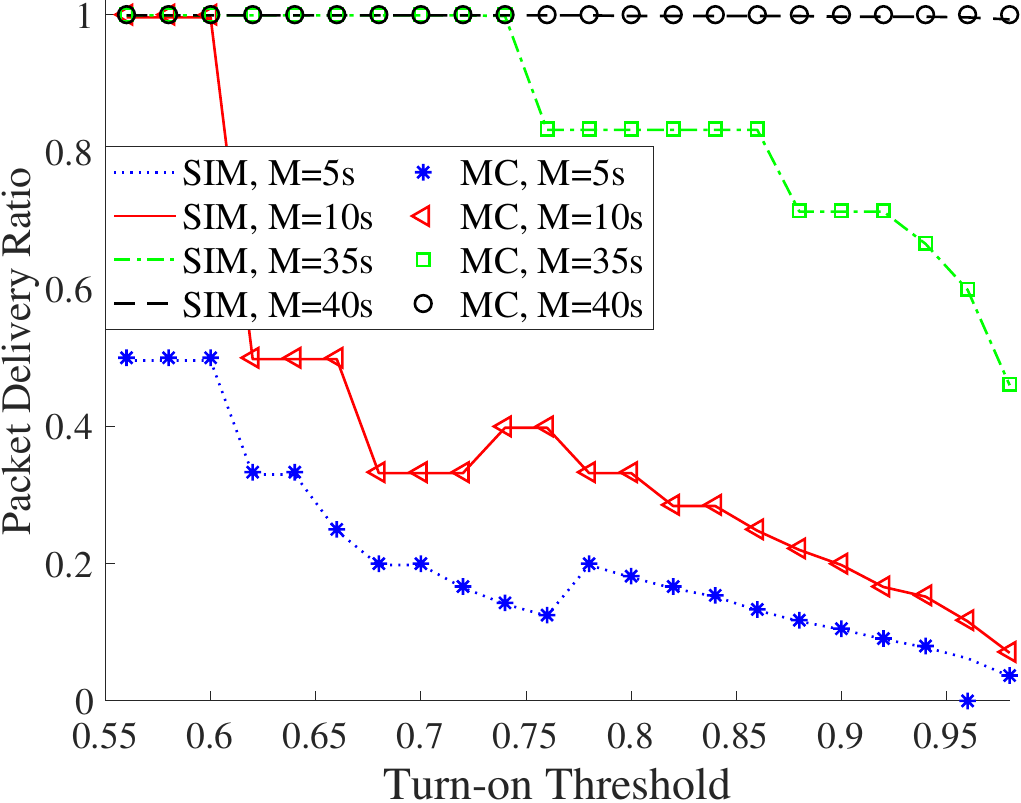}
			\label{fig:1dA_0_0_gran750}}
		
		\caption{PDR for different evaluations of case A ($SF=7$, $PS=8B$, $EH=0.001W$, $P_1=P_2=0$) of Table~\ref{tab:accuracyparam} when varying the turn-on threshold}
		\label{fig:ValidationCase1}
	}

	\end{figure*}

In this section, we validate the accuracy of the Markov Model by comparing it with the results of our simulator. As explained before, the simulation environment has been implemented as an event-based simulator in C++. For both the simulator and the Markov Model, the simulation setup is the same as the one presented in Section \ref{sec:simulationsetup}, but considering the ideal capacitor model.
The time unit of the Markov Model has been set to 1 second (i.e., $M$ is defined in seconds).

\subsection{Results and Discussion}\label{sec:accuracyresults}

In order to evaluate the accuracy of the Markov Chain model design, we compare its performance results with those obtained using the 
event-based simulation system in terms of the packet delivery ratio for the UL transmissions and for the DL receptions. The comparison has been made with the results obtained when averaging over 1000 transmissions for different scenarios, as it will be described later. The Markov Model is discrete and has a step size variable called $granularity$ which represents the accuracy of the rounded voltage levels. It changes the size of the matrix $P$. The finer the $granularity$ is, the larger the size of $P$ is. For example if  $granularity$ is 1, it would mean that the voltage is rounded to $1 V$ while if $granularity$ is 1000, it would mean that the voltage is rounded to $1 mV$. We are going to evaluate the accuracy of the model as a function of the $granularity$. 

\begin{table}[t!]
\begin{footnotesize}
\caption{Evaluated Scenarios}
\begin{center}
\begin{tabular}{c c c c c}
   \hline
\textbf{Case} &$\textbf{SF}$ & $\textbf{PS (Bytes)}$  & $\textbf{EH (W)}$ & \textbf{(\textbf{$P_1$}, \textbf{$P_2$})  }    \\
   \hline
 \textbf{A} & 7  & 8      & 0.001 & (0, 0), (1, 0) and (0, 1)     \\
 \textbf{B} &7  & 48     & 0.001 & (0, 0), (1, 0) and (0, 1)      \\
\textbf{C} & 9  & 48     & 0.01  & (0, 0), (1, 0) and (0, 1)     \\
 \textbf{D} &7  & 16    & 0.001 & (0, 0), (1, 0) and (0, 1)    \\
\textbf{E} & 9  & 16      & 0.001 & (0, 0), (1, 0) and (0, 1) \\
   \hline
\end{tabular}
\label{tab:accuracyparam}
\end{center}
\end{footnotesize}
\end{table}

For this evaluation, we have considered a capacitor of 4.7~mF and a Downlink Packet Size of 1 Byte (e.g., an ACK). Table~\ref{tab:accuracyparam} specifies the spreading factor ($SF$), the Uplink Packet Size (PS), the energy harvesting rate $P_{harvester}$ ($EH$) in Watts and the probabilities of downlink packets $P_1$ and $P_2$ for the 5 evaluated scenarios. 
Figure~\ref{fig:ValidationCase1} shows an example of case A of Table~\ref{tab:accuracyparam} where $P_1=P_2=0$ and the Packet Delivery Ratio is shown as a function of the turn-on threshold (which is represented as a percentage of the operating voltage~$E$). Figure~\ref{fig:1dA_0_0_gran100} shows the simulation results (SIM) compared with the Markov Chain Model (MC) for a $granularity$ value of 100 when the transmission interval $M$ takes the values of 5, 10, 35 and 40 seconds. Figures~\ref{fig:1dA_0_0_gran500} and \ref{fig:1dA_0_0_gran750} show the same results when the Markov Chain Model $granularity$ is set to 500 and 750 respectively. 

As can be seen,  Figure~\ref{fig:1dA_0_0_gran100} shows more differences between the Markov Chain model and the simulator than in Figures~\ref{fig:1dA_0_0_gran500} or \ref{fig:1dA_0_0_gran750} , which means that increasing the $granularity$, makes the error smaller. Even if the model is accurate in most cases, when the turn-on threshold is high ($>90\%$), the error increases, as can be seen in all the figures. 
At lower $granularities$, errors in PDR estimation start occurring at lower turn-on thresholds. 
When the $granularity$ is set to 750 (Figure \ref{fig:1dA_0_0_gran750}), only turn-on threshold values higher than 94\% have an absolute error appreciable, but it remains bellow 0.09. 
Besides, for the case of $M=5s$ (the blue line of Figures \ref{fig:1dA_0_0_gran100}, \ref{fig:1dA_0_0_gran500} and \ref{fig:1dA_0_0_gran750}), when the turn-on threshold is lower than 75\% of the operational voltage (0.75 in the X-axis of Figure~\ref{fig:ValidationCase1}), all the tested $granularity$ values perform accurately. However, when increasing the turn-on threshold, and depending on the $granularity$ value, the Markov Model becomes less accurate. When $granularity$ is set to 100, 
errors start appearing at turn-on thresholds of 76\% of the operational voltage (with an absolute error of 0.08 in terms of PDR), while some other turn-on thresholds provide more accurate results. If $granularity$ is set to 500 or 750, errors start appearing at turn-on thresholds of 88\%  or 96\%, respectively.

\begin{figure*}[t]
	\centering

	{
		\subfigure[Turn-on threshold of 0.70]{\includegraphics[width=0.66\columnwidth]{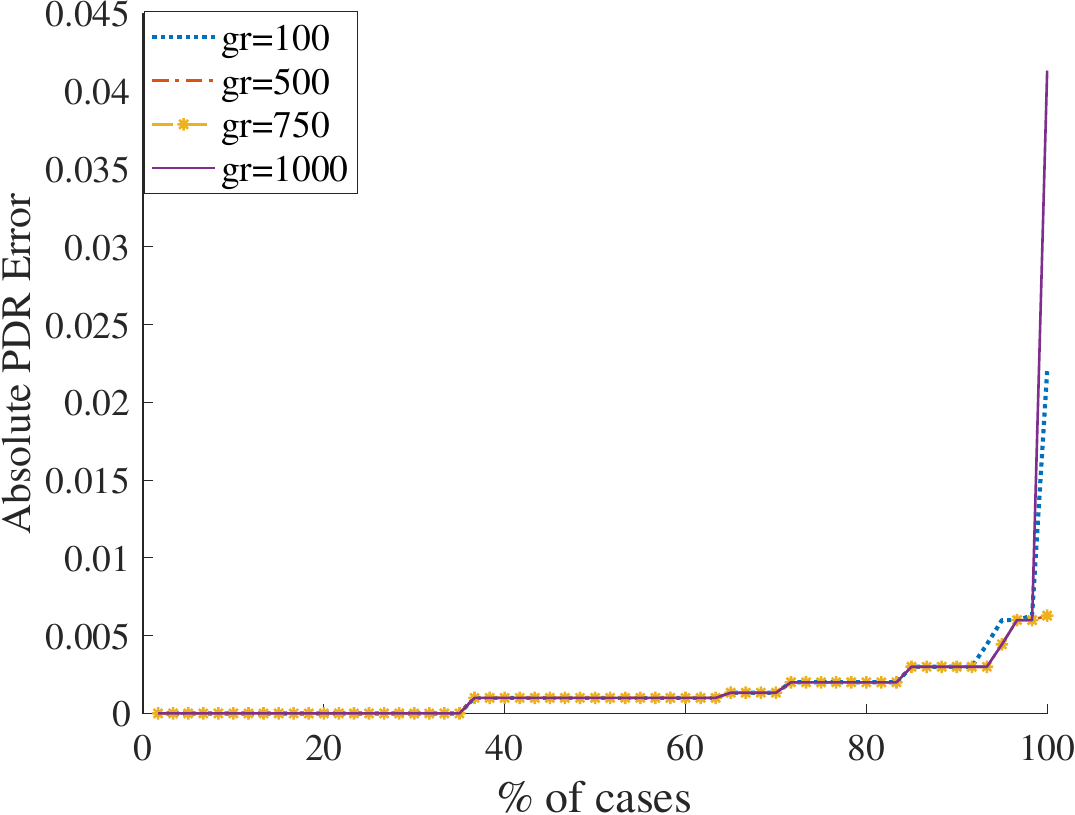}
			\label{fig:ULSortedErrorToT70}}
		\subfigure[Turn-on threshold of 0.84]{\includegraphics[width=0.66\columnwidth]{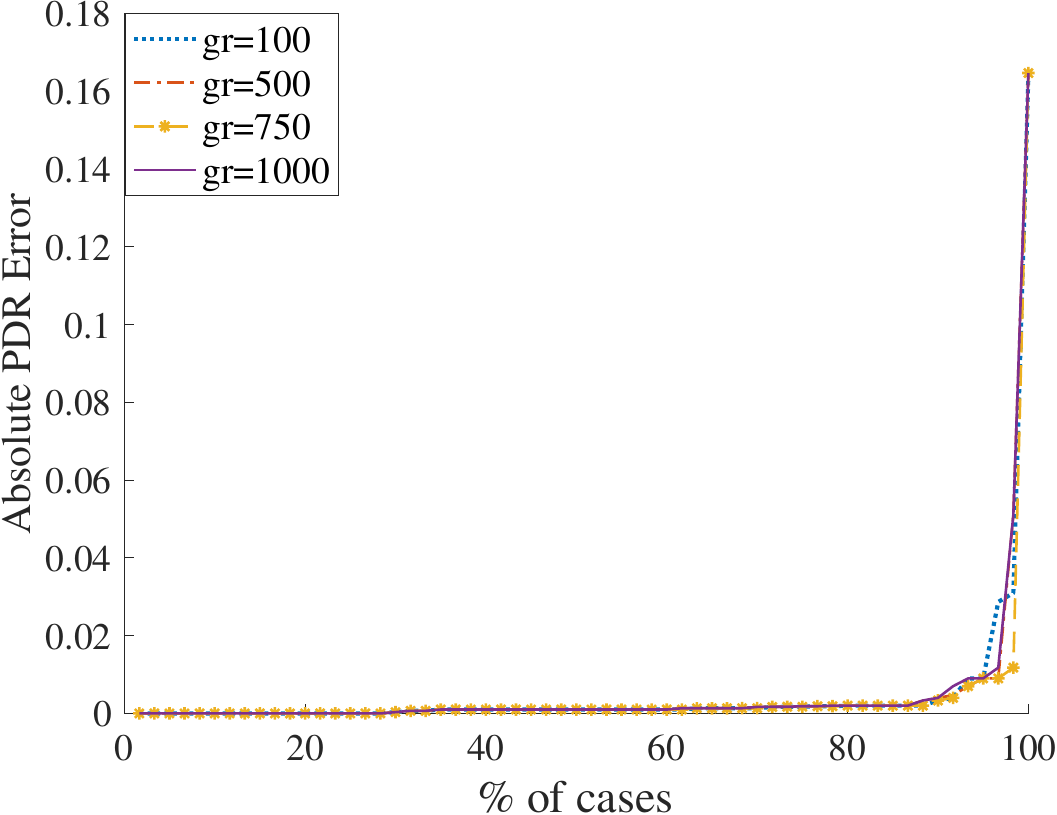}
			\label{fig:ULSortedErrorToT84}}
		\subfigure[Turn-on threshold of 0.96]{\includegraphics[width=0.66\columnwidth]{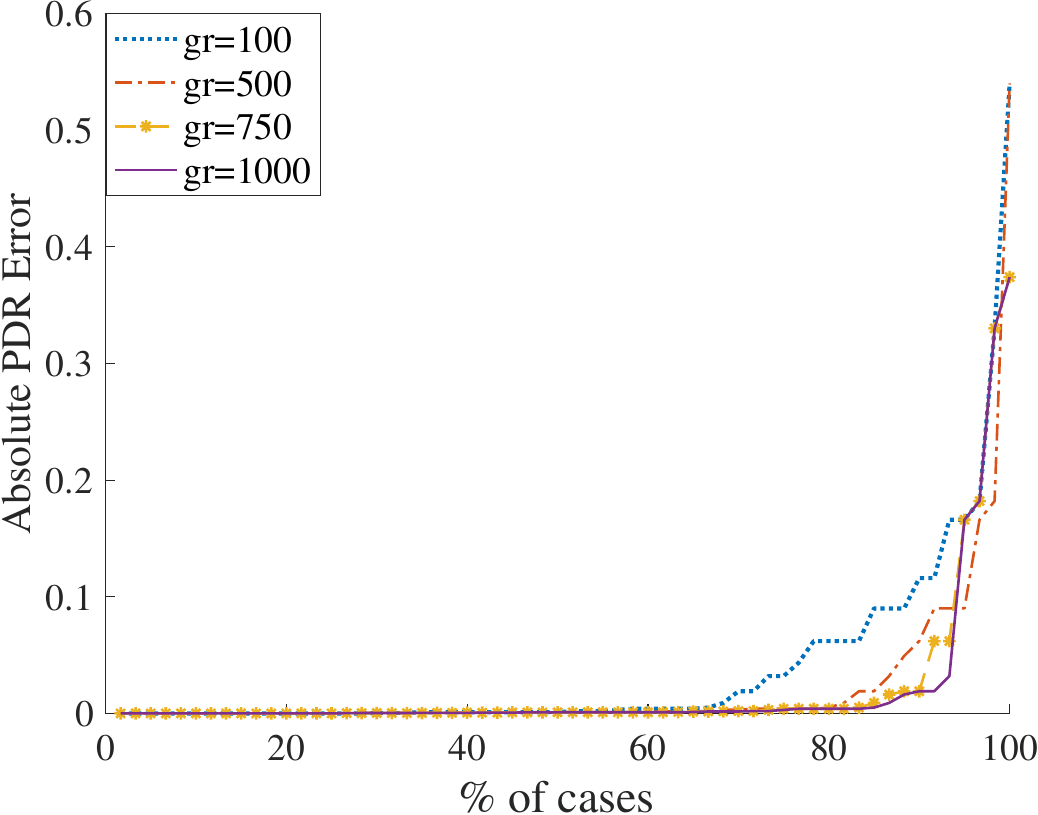}
			\label{fig:ULSortedErrorToT96}}
		
		\caption{Variation of the absolute UL PDR Error when varying the turn-on threshold}
		\label{fig:ValidationMC2}
	}

	\end{figure*}

In order to fully determine the optimal $granularity$, we have compared the complete set of cases of Table~\ref{tab:accuracyparam}. For every of those cases, we have considered four different transmission intervals: $small$ $M$, $medium$ $M$, $high$ $M$ and $very$ $high$ $M$. In the example of the Figure~\ref{fig:ValidationCase1} (case A), they correspond to 5, 10, 35 and 40 seconds respectively and are the same as for the case C, while for case B they correspond to 15, 20, 60 and 65 seconds, for case D they correspond to 5, 10, 40 and 45 seconds and for case E they correspond to 15, 30, 100 and 250 seconds respectively. In general, the value of $small$ $M$ has been chosen to consider the case where the maximum $PDR$ is 0.5. The value of $medium$ $M$ has been taken to guarantee a $PDR$ value of 1 for at least one value of the turn-on threshold. With $very$ $high$ $M$ we wanted to ensure that the $PDR$ was 1 for all the cases and $high$ $M$ is something in between.

Figure~\ref{fig:ValidationMC2} shows the absolute PDR error for UL transmissions of the Markov Model for all the cases cases of Table~\ref{tab:accuracyparam} when compared with the results of the simulator.  Figures~\ref{fig:ULSortedErrorToT70}, \ref{fig:ULSortedErrorToT84}  and  \ref{fig:ULSortedErrorToT96} show the results for a turn-on threshold value of 0.70, 0.84 and 0.96 respectively. Comparing the three figures, we can see the impact of the turn-on threshold in the absolute UL PDR error. 
While varying the value of $M$ makes no difference in the absolute PDR error, the higher the turn-on threshold is, the higher the error is. Besides, when setting the turn-on threshold to 0.70, for 90\% of all the cases for all the $granularity$ values, the error is negligible ($<$0.003). $Granularity$ values do not have much impact on smaller turn-on thresholds (even for 0.84, as can be seen in Figure~\ref{fig:ULSortedErrorToT84}). However, when setting the turn-on threshold to 0.96 (Figure~\ref{fig:ULSortedErrorToT96}), many differences are observed for the different $granularity$ values. While 90\% of the cases of granularity 1000 get an error below 0.02, only the 70\% of the cases of granularity 100 get the same performance. 

Finally, Table \ref{tab:compTime} shows the average and standard deviation execution time per scenario of the model. As can be seen, and as expected, the higher the $granularity$ is, the higher the CPU time is needed. For all the cases, $small$ $M$ needs more time to compute. This is due to the fact that more possibilities need to be evaluated in this scenario since not all the transmissions are successful due to the lack of voltage. 
In general terms, increasing the $granularity$ to values higher than 750 does not provide a statistically significant improvement in the accuracy while the execution time needed increases exponentially. For this reason, we will use a $granularity$ value of 750 in the remainder of the results.

\begin{table}[t]
\begin{scriptsize}

\caption{Averaged and standard deviation execution time per scenario (in seconds)}
\begin{center}

\begin{tabular}{l | c c c c}
\hline
&  \multicolumn{4}{c}{  \textbf{Granularity}  }\\
        & \textbf{100}               & \textbf{500}              & \textbf{750}              & \textbf{1000}             \\
  \hline
\textbf{\textit{small M}}   & 0.53s $\pm$ 0.028        	& 16.58s $\pm$ 1.30   & 56.7s $\pm$ 6.3   & 132.7s  $\pm$ 17     \\
\textbf{\textit{medium M} }   & 0.47s $\pm$ 0.032		& 13.11s $\pm$ 3.77		& 42.5s $\pm$ 14.9 & 96.4s $\pm$ 35.9 \\
\textbf{\textit{high M}}      & 0.44s  $\pm$  0.034    	& 9.9s  $\pm$    2.91    & 28.6s $\pm$ 12.3 & 65.8s $\pm$ 27.2 \\
\textbf{\textit{very high M}} & 0.43s $\pm$ 0.016 		& 8.32s $\pm$ 1.37 		& 21.7s $\pm$ 5.2   & 50.4s $\pm$ 11.4\\

\hline
\end{tabular}
\label{tab:compTime}
\end{center}
\end{scriptsize}
\end{table}

\section{Performance characterization of a Battery-less LoRaWAN Class A Device}\label{sec:Results}
In this section, we focus on the evaluation of the requirements in terms of the capacitor and the energy harvester, that will allow a battery-less LoRaWAN Class A device to work for different uplink and downlink transmission characteristics (determined by the packet size and the transmission interval), and for different environmental conditions.

We first provide an analysis of the needed capacitor depending on the conditions. 
The power consumed while performing one single LoRaWAN UL/DL cycle is usually higher than the power harvested by environmental harvesters. For this reason, in a cycle, the available energy decreases. That is why it is necessary to first calculate the minimum capacitance needed to perform an UL/DL cycle. 
And then, we evaluate how often these cycles can be performed, which would mean how often a LoRaWAN battery-less device can send data to the network. 
Finally, we analyse the impact of the turn-on voltage threshold 
on the reliability in terms of packet delivery ratio (PDR) for different data transmission rates. 
The simulation setup is the same as the one presented in Section \ref{sec:simulationsetup}.

\subsection{Analysis of the Capacitance}\label{sec:ResultsCapacitance}

The Class A LoRaWAN standard~\cite{LoraWan} defines the sequence to follow when performing a transmission as described in Section~\ref{sec:systemmodel}. In order to characterize the system, it would be useful to know what kind of capacitor is needed to support it (i.e., that is able to save enough harvested energy to perform the needed tasks), since it is one of the main components in a battery-less device. 
The needed capacitance to perform a single UL/DL cycle will depend on the UL and DL characteristics, and on the specific system parameters shown in Tables~\ref{tab:simStates}-\ref{tab:generalparameters}. 

\begin{figure}[t]
\centerline{\includegraphics[width=.9\columnwidth]{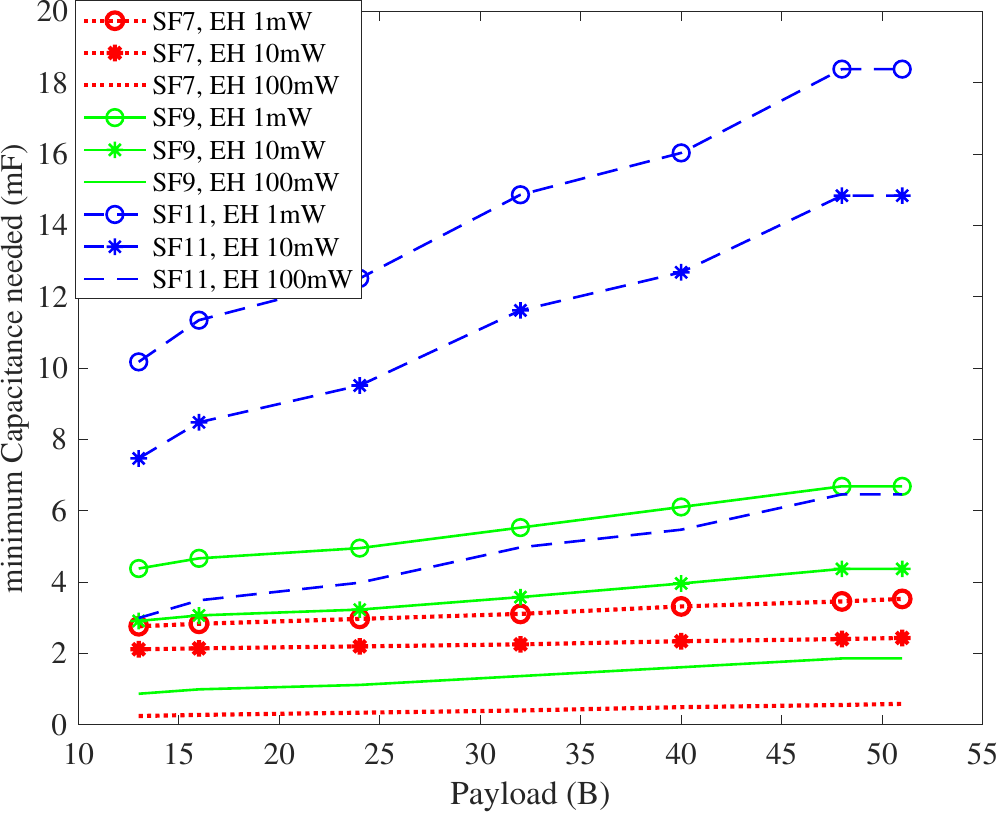}}
\caption{Minimum capacitance needed when varying the payload for UL packets when no DL is received}
\label{fig:minC_p1_0p2_0}
\end{figure}

\subsubsection{Only UL}

Figure~\ref{fig:minC_p1_0p2_0} shows the minimum capacitance ($C$) needed to complete one UL transmission cycle for different values of $SF$, and energy harvesting when varying the payload for UL packets. Such a cycle includes transmitting a packet and staying awake for RX1 and RX2 (assuming $P_1$ and $P_2$ are zero). These results have been obtained for the specific values shown in Tables \ref{tab:simStates} and \ref{tab:generalparameters}.

For high data rates (i.e., small $SF$ values), the payload has little influence on the required capacitance. However, when using $SF11$, the impact of the packet size is considerable (e.g., varying from $10.2mF$ to $18.4mF$ when the energy harvesting rate is set to $1mW$). 
As expected, lower energy harvesting rates are only compatible with larger spreading factors if supercapacitors are used, which could not always be compatible with the low cost and small form factor requirements of IoT devices.

\begin{figure*}[t]
	\centering
	{
		\subfigure[DL packet size = 1B]{\includegraphics[width=0.65\columnwidth]{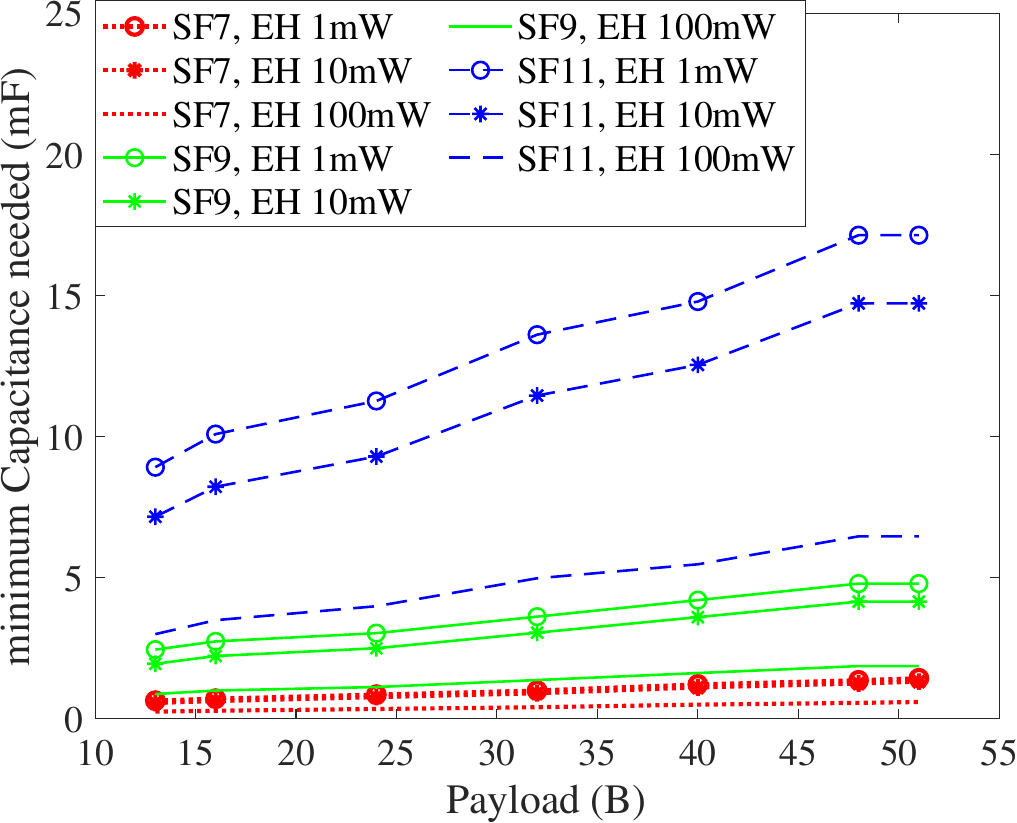}
			\label{fig:minC_p1_1p2_0_dlps1}}
			\subfigure[DL packet size = 16B]{\includegraphics[width=0.65\columnwidth]{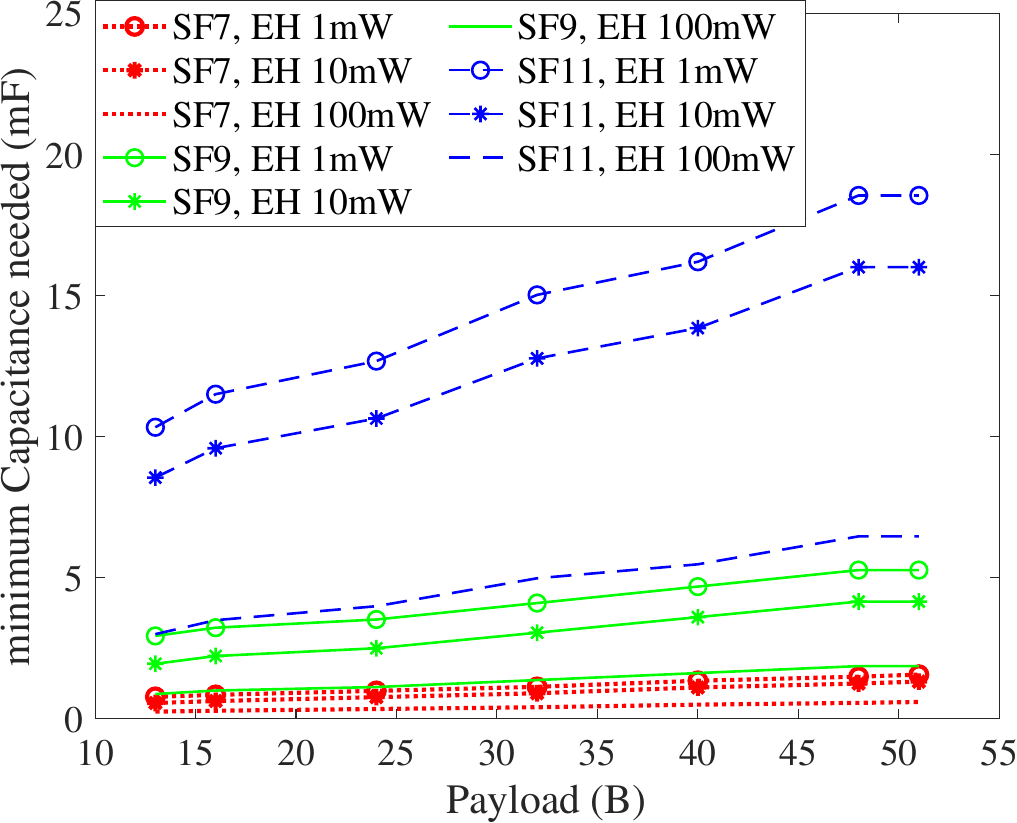}
			\label{fig:minC_p1_1p2_0_dlps16}}
		\subfigure[DL packet size = 48B]{\includegraphics[width=0.65\columnwidth]{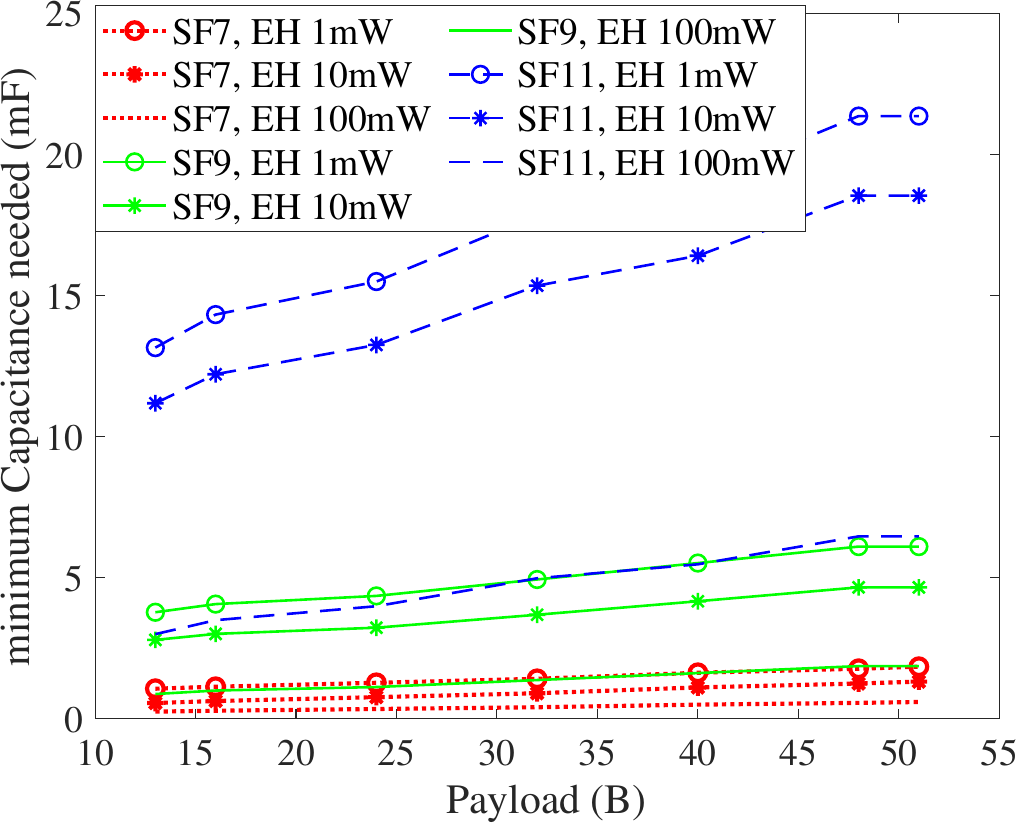}
			\label{fig:minC_p1_1p2_0_dlps48}}
		\caption{Minimum capacitance needed for UL and DL in the first window when varying the payload for UL packets}
		\label{fig:minC_p1_1p2_0}
	}
\end{figure*}

\begin{figure*}[t]
	\centering
	{
		\subfigure[DL packet size = 1B]{\includegraphics[width=0.65\columnwidth]{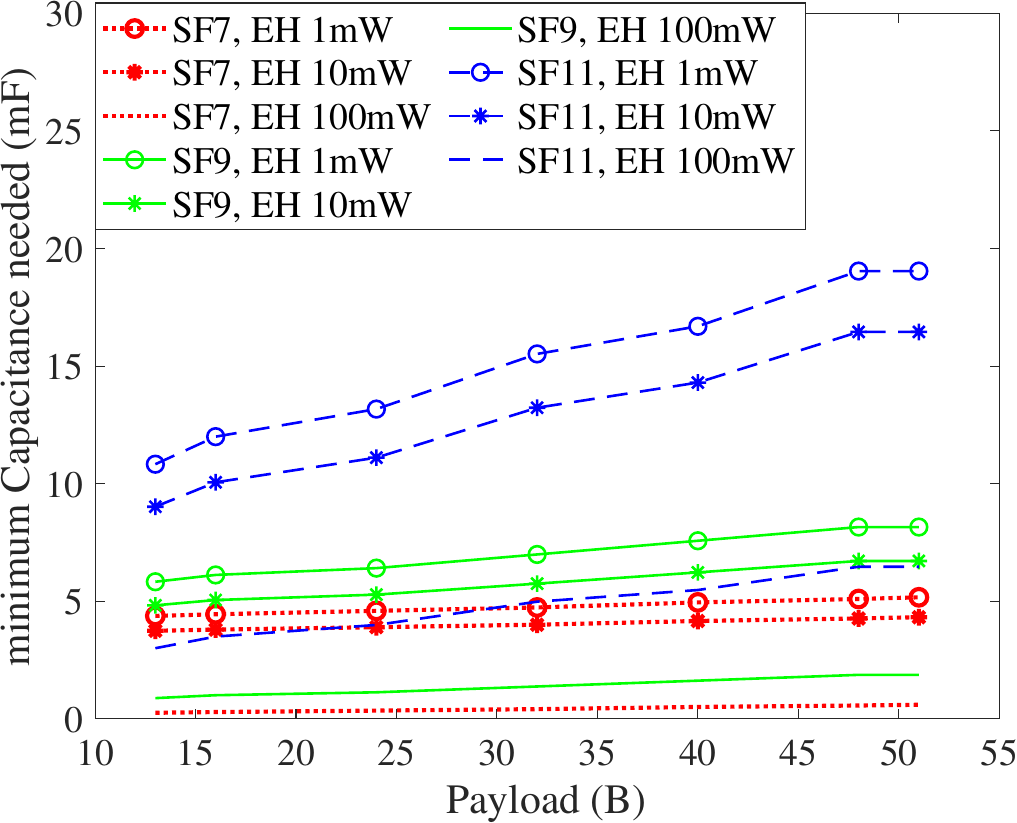}
			\label{fig:minC_p1_0p2_1_dlps1}}
		\subfigure[DL packet size = 16B]{\includegraphics[width=0.65\columnwidth]{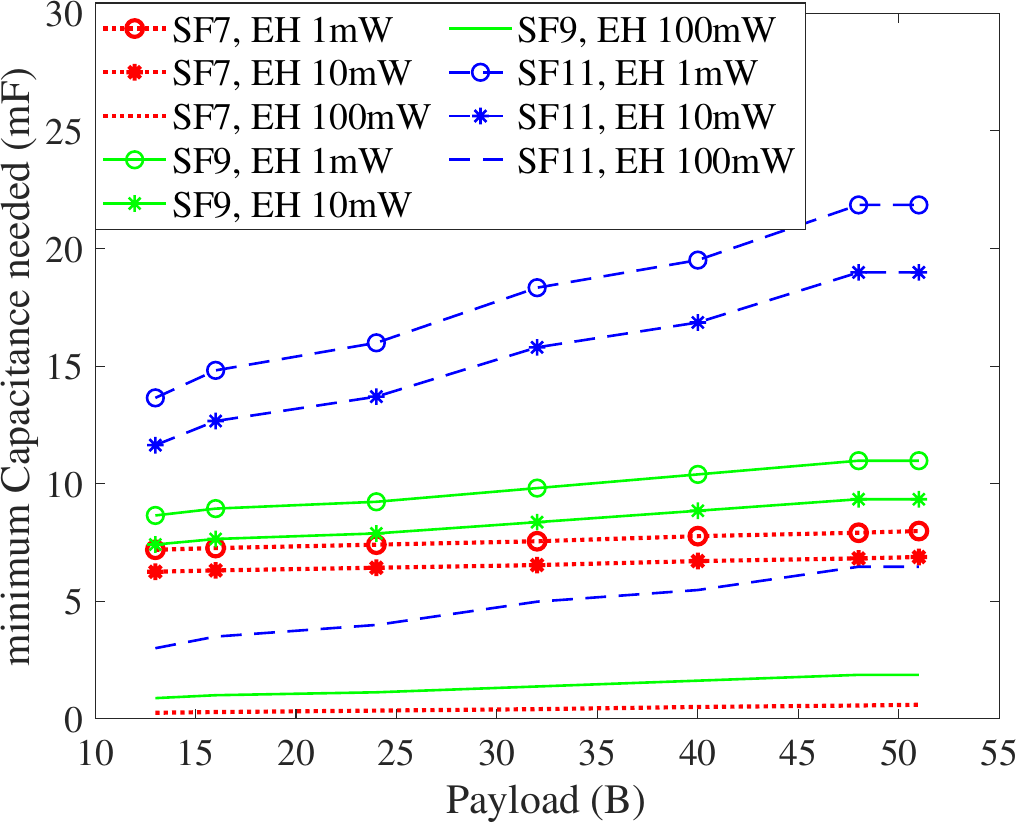}
			\label{fig:minC_p1_0p2_1_dlps16}}
		\subfigure[DL packet size = 48B]{\includegraphics[width=0.65\columnwidth]{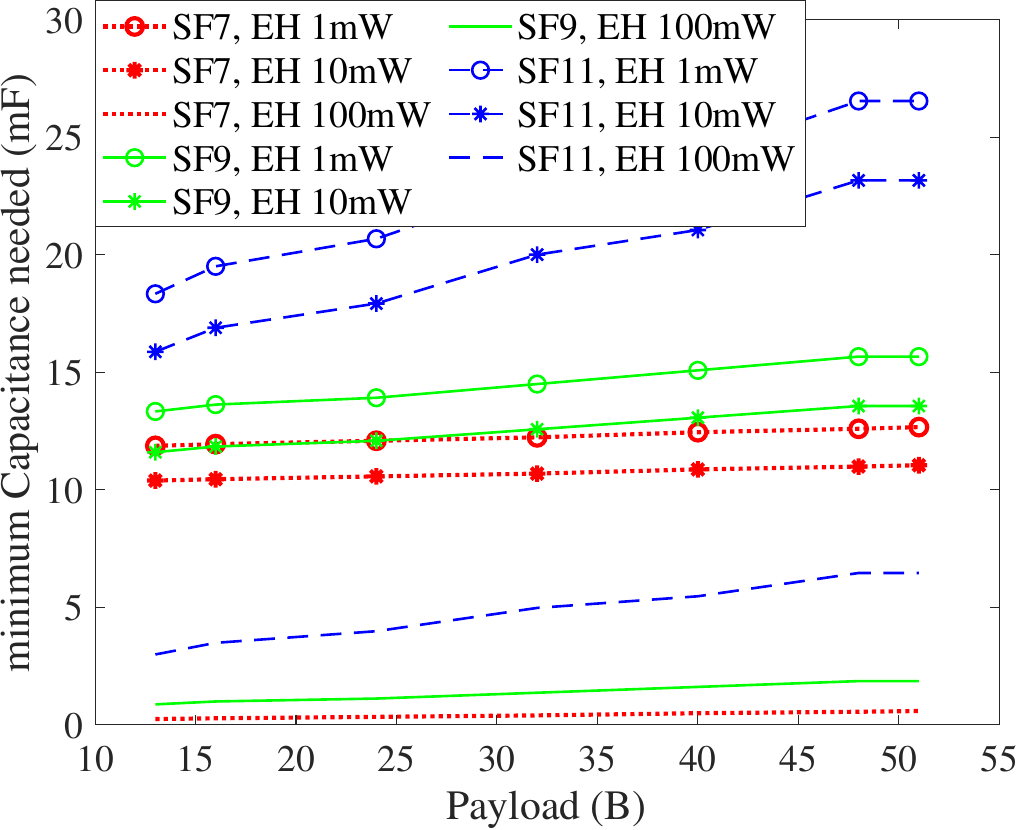}
			\label{fig:minC_p1_0p2_1_dlps48}}
		\caption{Minimum capacitance needed for UL and DL in the second window when varying the payload for UL packets}
		\label{fig:minC_p1_0p2_1}
	}
\end{figure*}

\subsubsection{UL and DL}

Figure~\ref{fig:minC_p1_1p2_0} shows the minimum capacitance needed to complete one UL transmission cycle followed by a DL reception in the first window (i.e. $P_1 = 1$) when varying the payload for UL packets for different values of DL packet size: 1 Byte in Figure~\ref{fig:minC_p1_1p2_0_dlps1}, 16 Bytes in Figure~\ref{fig:minC_p1_1p2_0_dlps16} and 48 Bytes in Figure~\ref{fig:minC_p1_1p2_0_dlps48}. 

When using high data rates (i.e., $SF7$), there is no impact on the capacitance needed while varying the UL nor DL packet size. However, for high $SF$ values, the capacitance needed increases when increasing the packet size of both UL and DL. In Figure~\ref{fig:minC_p1_1p2_0_dlps16}, we can see that if the DL packet size is fixed to 16 Bytes and using 1mW of energy harvesting and $SF11$, the needed capacitance for UL packet sizes of 16 Bytes and 48 Bytes, varies from $11.5mF$ to $18.55mF$. And when the UL packet size is fixed to 16 Bytes and we vary the DL packet size from 16 Bytes to 48 Bytes, the needed capacitance varies from $11.5mF$ to $14.3mF$ (Figures~\ref{fig:minC_p1_1p2_0_dlps16} and \ref{fig:minC_p1_1p2_0_dlps48}). This means that the UL packet size has more impact than DL packet size, due to the fact that the power consumption for transmitting is higher than for receiving for the radio used (the Semtech SX1272/73 LoRa radio \cite{Semtech}).

When the DL is received in the second window, the differences are more important.  
Figure~\ref{fig:minC_p1_0p2_1} shows the minimum capacitance needed for UL and DL in the second window (i.e. $P_1 = 0$ and $P_2 = 1$) when varying the payload for UL packets for different values of DL packet size: 1 Byte in Figure~\ref{fig:minC_p1_0p2_1_dlps1}, 16 Bytes in Figure~\ref{fig:minC_p1_0p2_1_dlps16} and 48 Bytes in Figure~\ref{fig:minC_p1_0p2_1_dlps48}. 

When a downlink packet is received in the second window, a bigger capacitance ($C$) is needed to successfully complete the cycle. This is due to the fact that $SF12$ is always used in this window, so that the slowest data rate is employed. When using high values of $EH$, DL packet size does not impact the needed capacitance since the Idle periods between the UL and DL are enough to harvest the needed energy for the next reception attempt (i.e. during the second Idle period the device can harvest the needed amount of energy to receive the DL packet with $SF12$). However, if a smaller $EH$ rate is considered (1~mW), these Idle periods are not enough to recharge the energy and bigger capacitances are needed.

To sum up, if no DL is received, a capacitance of $3.5mF$ is sufficient to support transmissions using $SF7$ for all the $EH$ cases, $6.7 mF$ are needed for $SF9$, and the needed capacitance for $SF11$ is $18.3 mF$. If a 1 Byte downlink is received in the first window, the needed capacitance is reduced to $1.4 mF$ for $SF7$ and to $17.2 mF$ for $SF11$. However, if the downlink packet size increases to 48 Bytes, the needed capacitance varies from $1.9 mF$ for $SF7$ up to $21.4 mF$ for $SF11$.
Finally, the worst case happens when the downlink packet is received in the second window, and for this case, if the DL packet size is 48 Bytes, a $13 mF$ capacitor is needed to support transmissions with $SF7$ for all the shown $EH$ cases, and $16 mF$ and $27 mF$ will allow $SF9$ and $SF11$ transmissions, respectively.

\begin{figure*}[t]
	\centering
	{
		\subfigure[No DL]{\includegraphics[width=0.65\columnwidth]{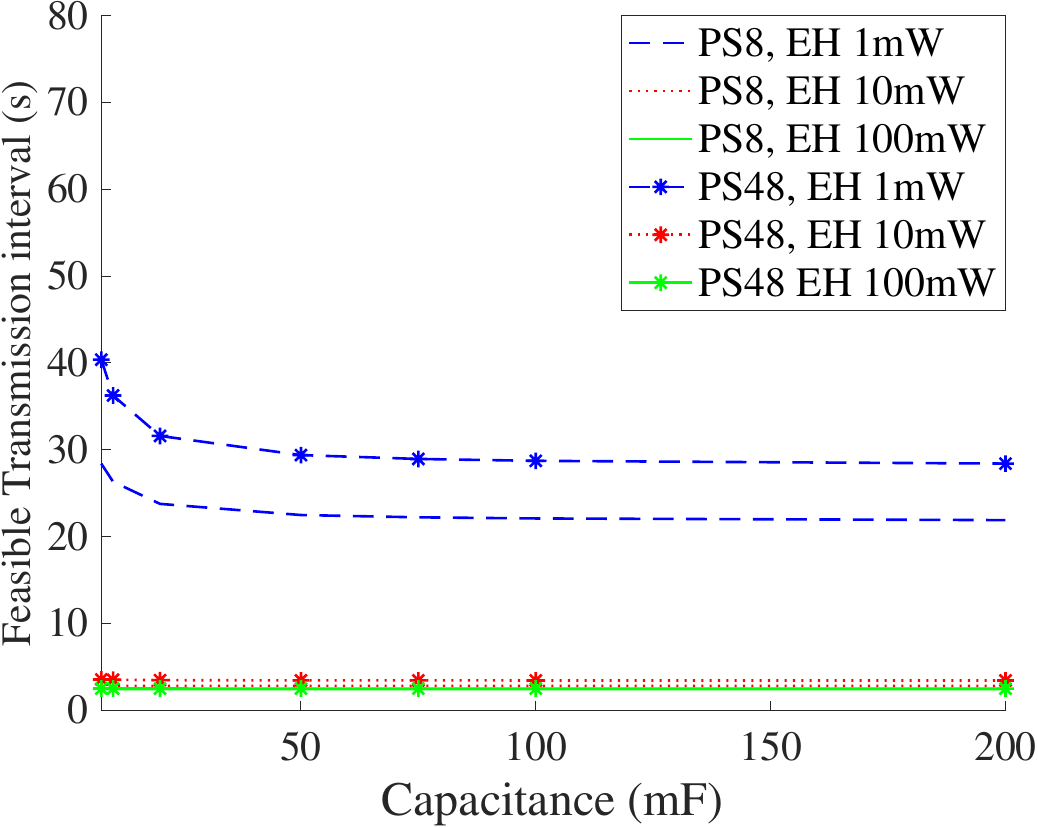}
			\label{fig:SF7_TxInterval_P1=0_P2=0}}
		\subfigure[DL in RX window 1]{\includegraphics[width=0.65\columnwidth]{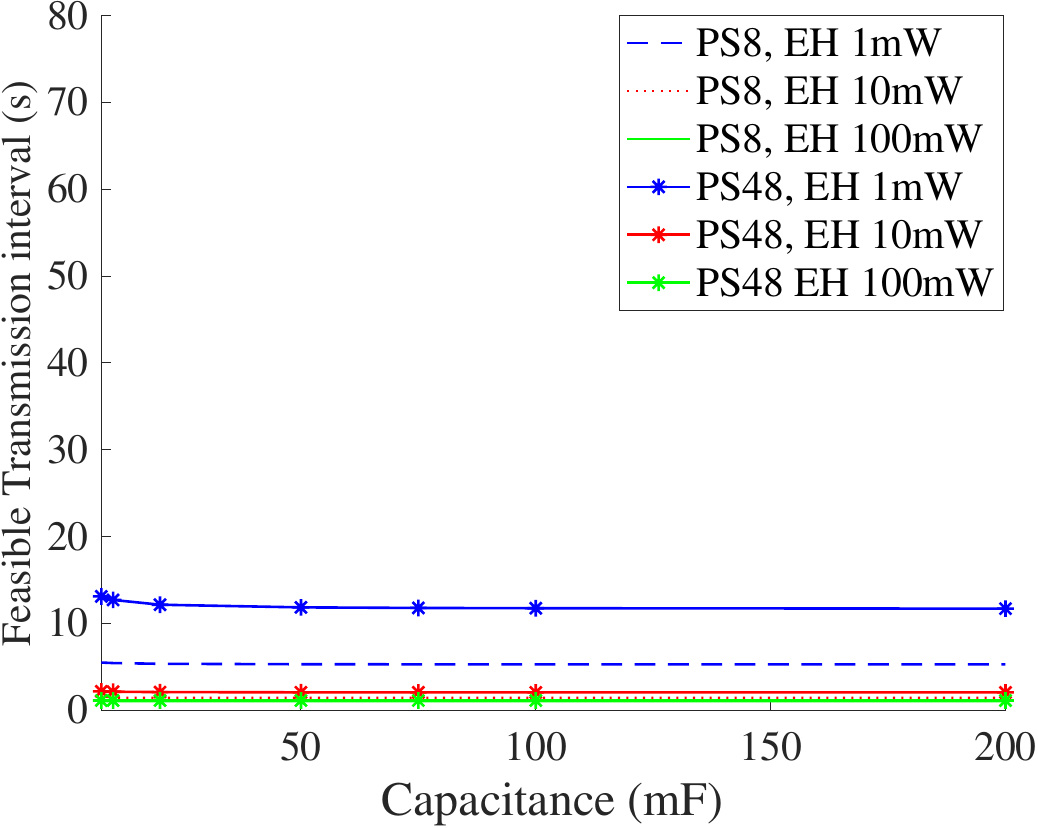}
			\label{fig:SF7_TxInterval_P1=1_P2=0}}
		\subfigure[DL  in RX window 2]{\includegraphics[width=0.65\columnwidth]{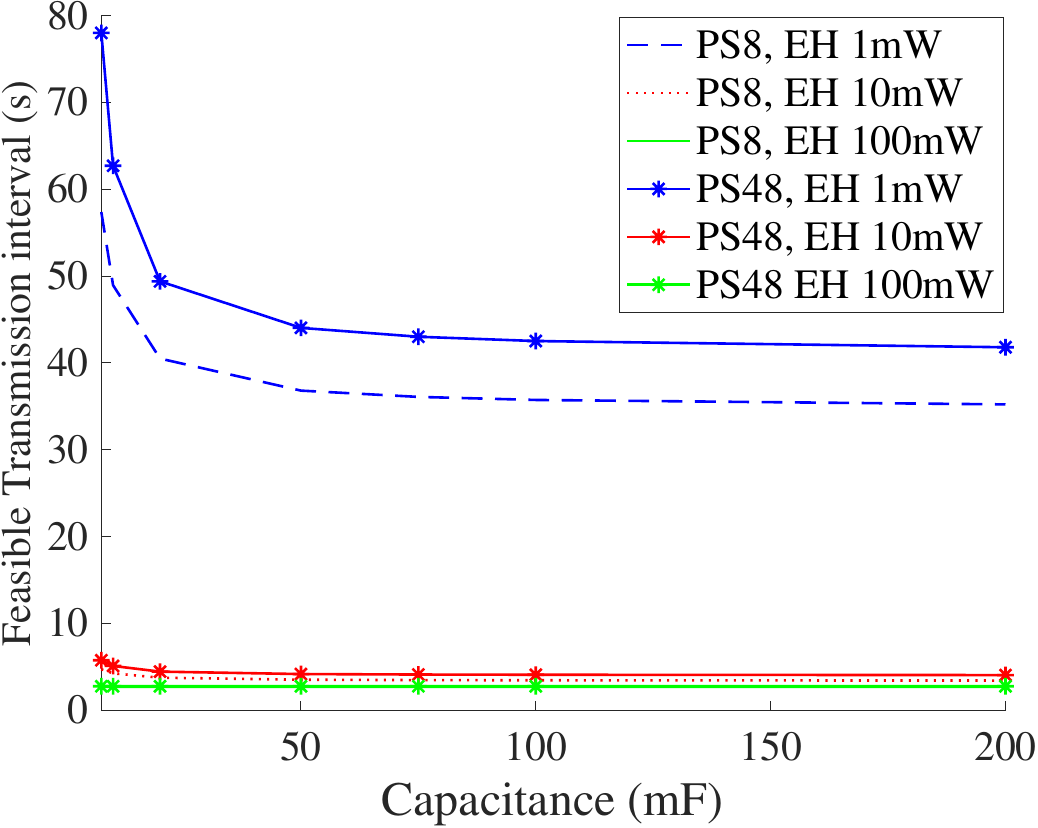}
			\label{fig:SF7_TxInterval_P1=0_P2=1}}
		\caption{Minimum Transmission Interval considering the device just wakes up to transmit and goes to off directly for $SF7$ and DL PS of 1B}
		\label{fig:SF7_TxInterval}
	}
\end{figure*}

\begin{figure*}[t]
	\centering
	{
		\subfigure[No DL]{\includegraphics[width=0.65\columnwidth]{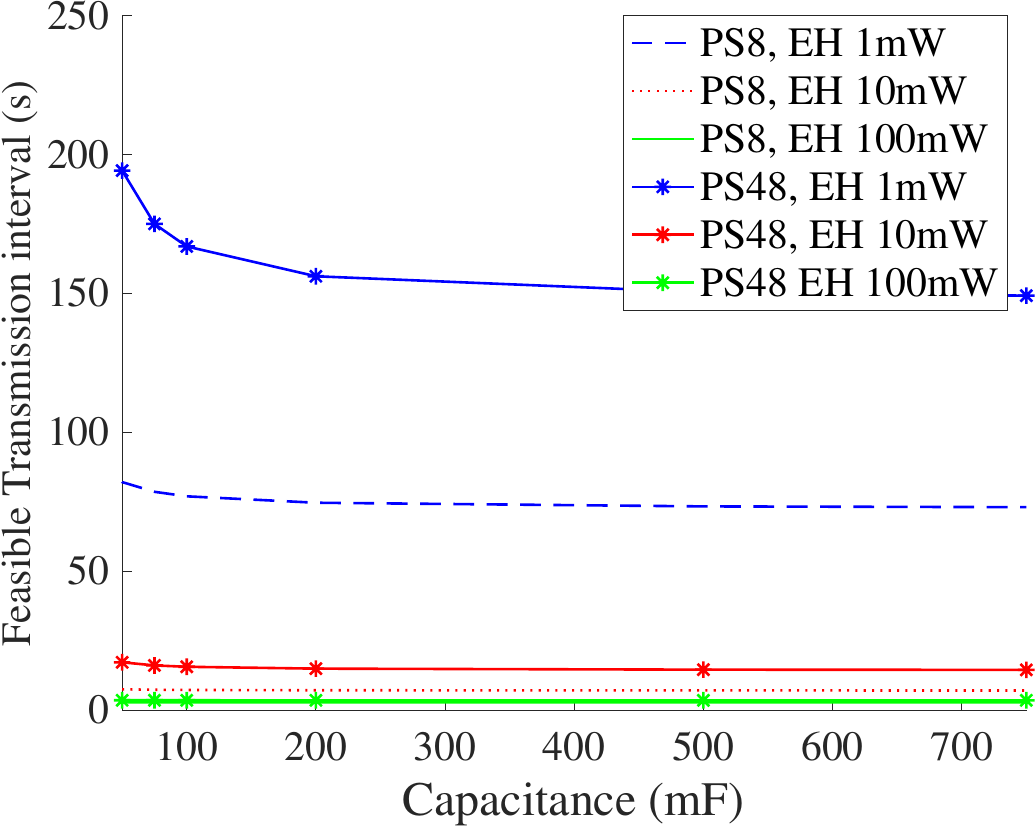}
			\label{fig:SF11_TxInterval_P1=0_P2=0}}
		\subfigure[DL in RX window 1]{\includegraphics[width=0.65\columnwidth]{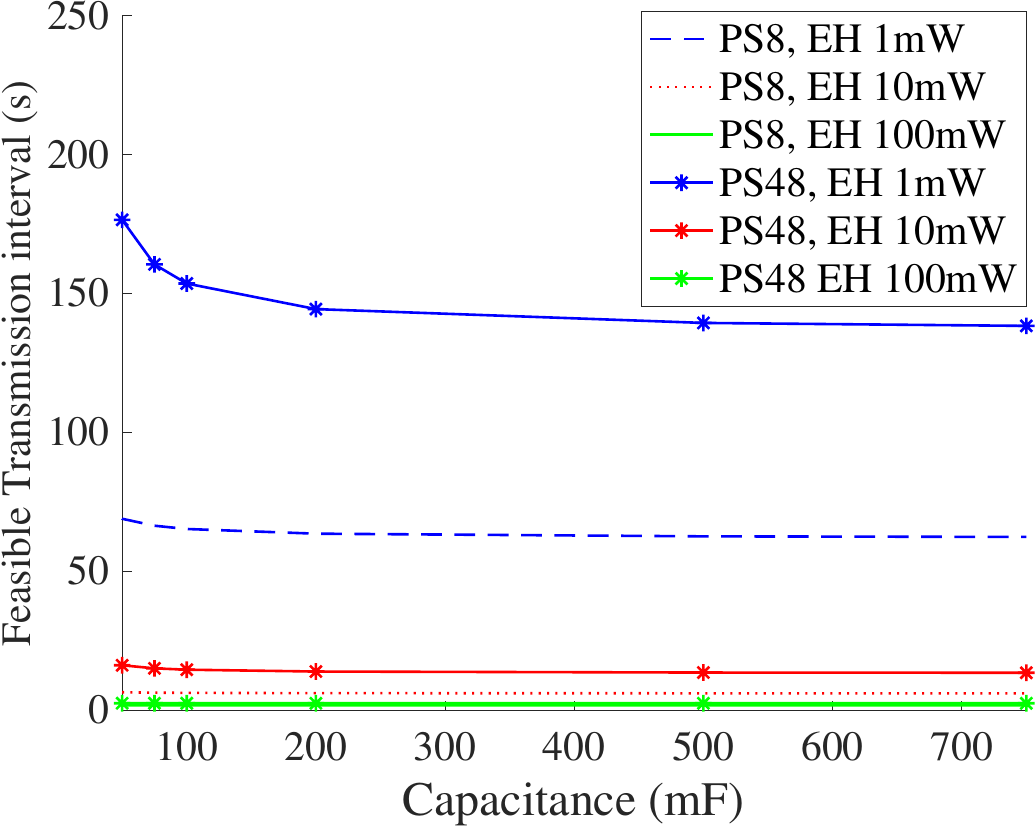}
			\label{fig:SF11_TxInterval_P1=1_P2=0}}
		\subfigure[DL in RX window 2]{\includegraphics[width=0.65\columnwidth]{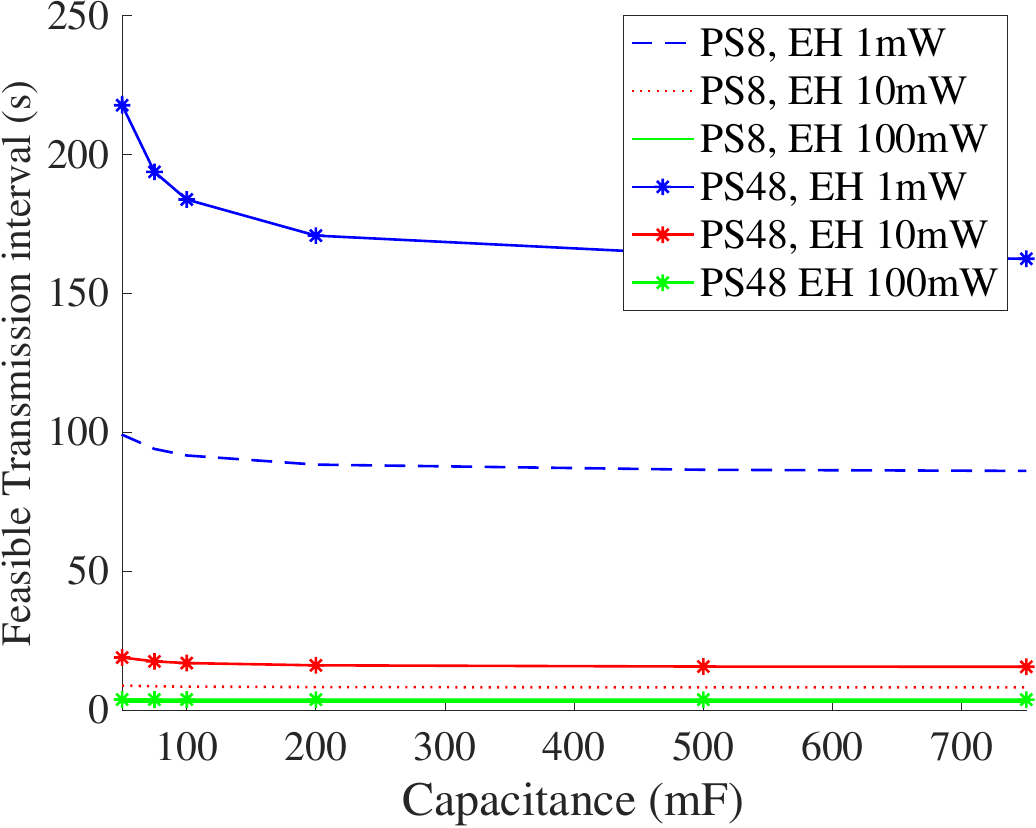}
			\label{fig:SF11_TxInterval_P1=0_P2=1}}
		\caption{Minimum Transmission Interval considering the device just wakes up to transmit and goes to off directly for $SF11$ and DL PS of 1B}
		\label{fig:SF11_TxInterval}
	}
\end{figure*}

\subsection{The feasible Transmission Interval}\label{sec:ResultsInterval}

The feasible transmission interval can be defined as the minimum transmission interval that a battery-less LoRaWAN device can successfully achieve. According to the specific requirements of the transmission ($SF$, PS or DL PS), the capacitor will need to be charged with at least some amount of energy to be able to perform the UL/DL cycle, so there will be a required time to charge the capacitor from $V_{min}$ up to the values needed to perform such a cycle.
This analysis shows the fastest transmission interval achievable for a battery-less  Class A LoRaWAN device, which will imply that just after completing the cycle, the device runs out of battery and turns-off.

Receiving a packet in the second window always requires more energy (i.e. more time to charge the capacitor) since it needs to listen during the first window and then receive the packet with $SF12$ in the second window. It is interesting to note that receiving a 1 Byte packet in the first window requires less energy (i.e. less time to charge the capacitor) than no DL due to the fact that if nothing is received in the first window, then the device has to listen for the preamble (with $SF12$) during the second window. 

Figure~\ref{fig:SF7_TxInterval} shows the results for a transmission with $SF7$ and a DL packet size of 1 Byte. As expected, more time is needed for the case where DL is received in the second window (Figure~\ref{fig:SF7_TxInterval_P1=0_P2=1}), followed by the case where no DL is received (Figure~\ref{fig:SF7_TxInterval_P1=0_P2=0}). Finally, faster transmission intervals can be achieved if receiving a DL of 1 Byte in the first window (Figure~\ref{fig:SF7_TxInterval_P1=1_P2=0}). For example, Figure~\ref{fig:SF7_TxInterval_P1=0_P2=1} shows that using a capacitance of 20 mF with an energy harvester of 1 mW, the device could send a 48 Bytes packet every 50 seconds if a downlink of 1 Byte is received in the second window, while if the downlink is not received, this time can be shortened to 32~s (Figure \ref{fig:SF7_TxInterval_P1=0_P2=0}). As can be seen, for all the cases, no benefit is achieved when increasing the capacitor size to more than 100 mF. This is related 
to the fact that there is always some energy required to wake up the device and this means that it will need some time to charge the capacitor to this point. This is a threshold in the minimum time a device can perform the UL/DL cycle that will depend on the harvested energy and the specific wake up energy needed.

Figure~\ref{fig:SF11_TxInterval} shows the results for a transmission with $SF11$. Compared with the values of using $SF7$, more time is needed between two transmissions since more voltage is required for the UL (and also DL in the first window). Figure~\ref{fig:SF11_TxInterval_P1=0_P2=1} has the highest times as expected, which vary from 217 s ($C=50mF$) to 163 s ($C=500 mF$) for the worst case where PS is set to 48 Bytes and $P_{harvester} = 1mW$. 
Again, no DL (Figure~\ref{fig:SF11_TxInterval_P1=0_P2=0})  and DL using the first window (Figure~\ref{fig:SF11_TxInterval_P1=1_P2=0}) show the best results when  $SF11$ is used. In this case, no benefit is achieved when increasing the capacitor size to more than 200 mF, which is again related to the minimum time required to wake up.

In summary, given high harvesting power, transmissions can be done at least every 3 seconds if using $SF7$, no matter if a downlink will be received or not. However, in case of 1~mW of harvesting power, the capacitor used influences the achievable transmission rate, which varies from 78 seconds to 43 seconds for a capacitance of $7.5 mF$ or $100 mF$ when a UL packet size of 48~Bytes is sent and 1~Byte packet size is received in the second window, or from 58~s ($C = 7.5 mF$)  to 36 s ($C = 100 mF$) when reducing the UL packet size to 8~Bytes. 
On the other side, when using $SF11$, if the harvesting power is high, 
transmissions can be done at least every 4 seconds, no matter if a downlink will be received or not. However, in case of 1 mW of harvesting power, the capacitor used also influences the achievable transmission rate, which varies from 218 s to 171 s for a capacitance of $50 mF$ or $200 mF$ when the UL packet size is set to 48 Bytes and 1 Byte packet size is received in the second window, or from 100 s ($ C = 50 mF$)  to 89 s ($C = 200 mF$) for a UL packet size of 8 Bytes.

\subsection{Analysis of the optimal turn-on threshold}\label{sec:ResultsTurnOnThreshold}

In the following we consider three different capacitors to evaluate the feasibility of the design. Let us consider first a capacitor with $C = 4.7mF$, which is in line with the capacitance of low-cost off-the-shelf Aluminum-Polymer capacitors and will only allow transmissions using $SF7$ or $SF9$. Second, we consider a supercapacitor (which can be used for all the cases shown in Figures~\ref{fig:minC_p1_0p2_0}-\ref{fig:minC_p1_0p2_1}) of $C = 47m{}F$ (such as the Kemet FM0V473ZF), which will be able to support $SF11$ and finally a bigger supercapacitor with $C = 1F$ (such as the Panasonic EECRG0V105V), to better see the impact when lower energy harvesting rates are used.

Before we have analysed the best case for the minimum transmission interval if the device just wakes up when the needed voltage is enough for completing a cycle. However, most of the times, it is unknown if a DL will take place, and what the DL packet size will be, which means that the device can not wake up in an intelligent way assuming what is the needed voltage to perform the cycle. For this reason, using the values previously shown might not be realistic, and we need to introduce a fixed (but configurable) turn-on voltage threshold at which the device will wake up. In the following, we represent this turn-on voltage threshold as a percentage of the maximum voltage of the system ($E$). 
Besides, although in Figures~\ref{fig:SF7_TxInterval}-\ref{fig:SF11_TxInterval} it was shown that the bigger the capacitor is, the faster the transmission interval that can be achieved, using a bigger capacitor will require much more time to wake up.

\begin{figure}[t]
\centerline{\includegraphics[width=0.8\columnwidth]{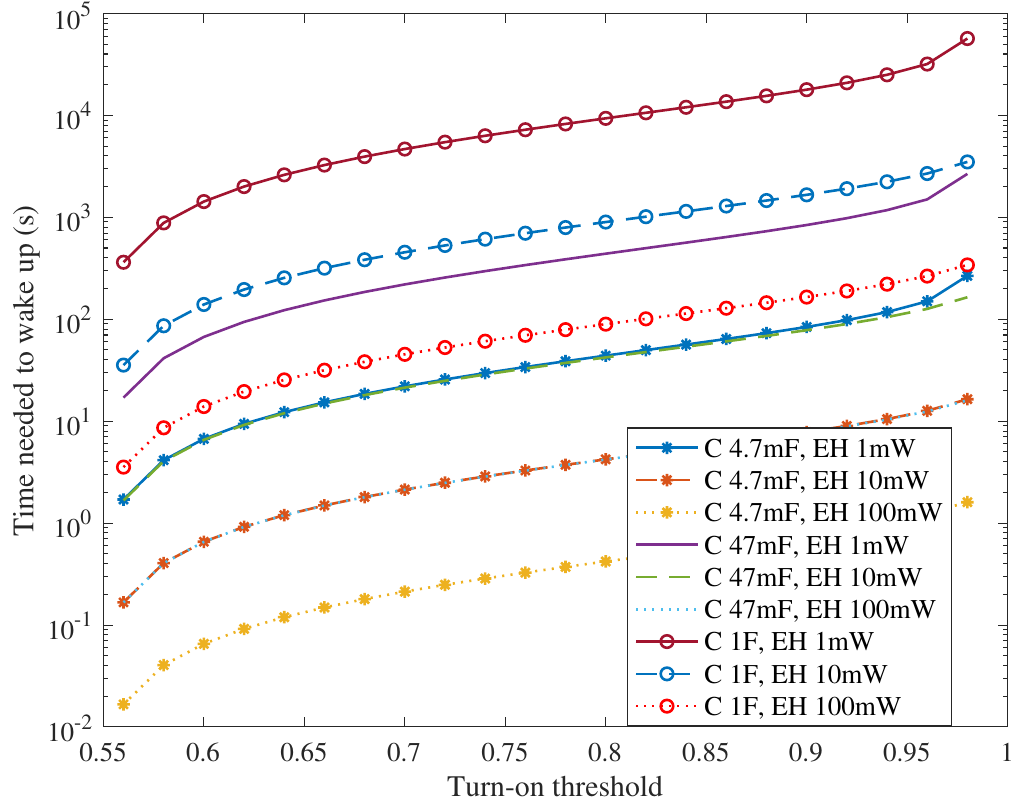}}
\caption{Needed time for waking up for different turn-on thresholds}
\label{fig:TimeToWU}
\end{figure}

Figure~\ref{fig:TimeToWU} shows the time needed to turn-on when the device is in the Off state and starting from $V_{min}$ for the three capacitors presented and for different values of $P_{harvester}$. 
We vary the turn-on threshold from 55\% (i.e., slightly higher than the 1.8~V turn-off threshold) up to 98\%. As can be seen, the minimum time needed is achieved for low turn-on thresholds, and also when increasing the energy harvesting power ($P_{harvester}$). For the smallest capacitor (4.7 mF) with the best energy harvester of 100 mW, if the turn-on threshold is 56\%, it can wake up within 0.017 s, while for the supercapacitor of 1~F, this will take 3.55 s. These differences will impact both the PDR and also the transmission interval rate. This also gives us the reason why it is important to choose the proper capacitor size according to the specific environmental conditions (energy harvesting rate) and network conditions ($SF$ and payload size), as well as intelligent turn-on mechanisms that take into account those conditions.

Since the turn-on threshold influences the time to wake up, in this section we evaluate how the turn-on threshold affects the performance of the system, varying it from 55\%  (i.e., slightly higher than the 1.8V turn-off threshold) 
up to 98\%. We evaluate it in different situations (only UL, with DL in the first window and with DL in the second window) and for the three different capacitor sizes and different transmission rates.

\begin{figure*}[t]
	\centering
	{
		\subfigure[No DL, only Tx PDR ]{\includegraphics[width=0.65\columnwidth]{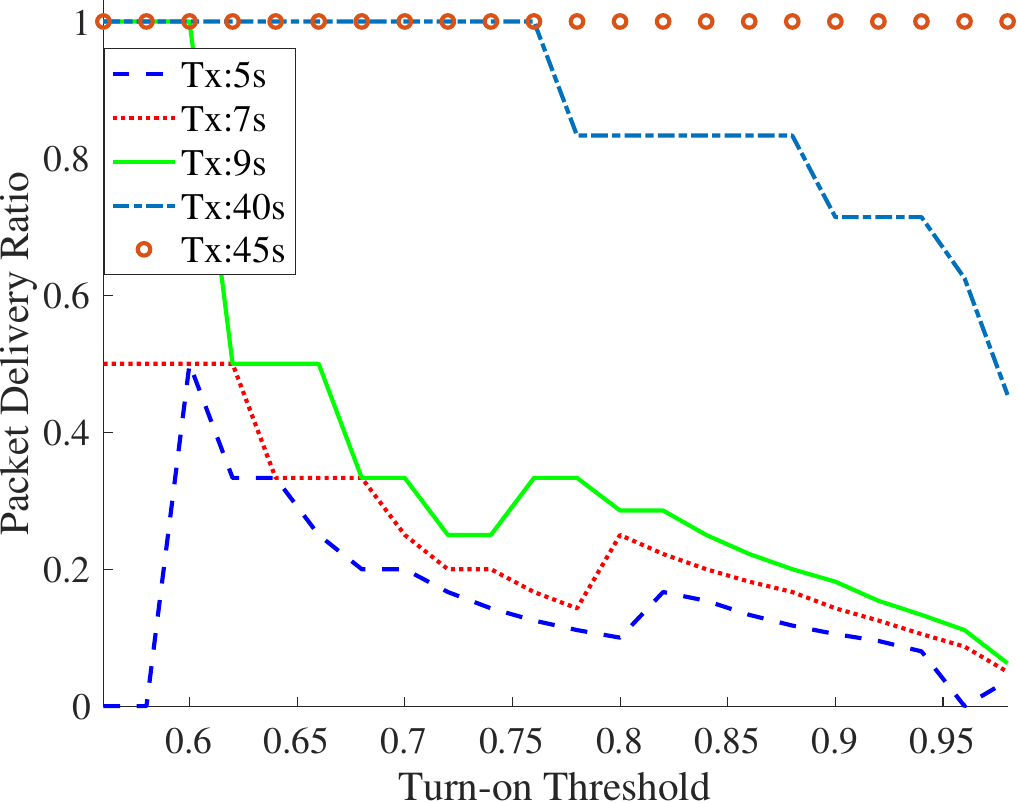}
			\label{fig:C1_NoDL}}
		\subfigure[DL in RX window 1]{\includegraphics[width=0.65\columnwidth]{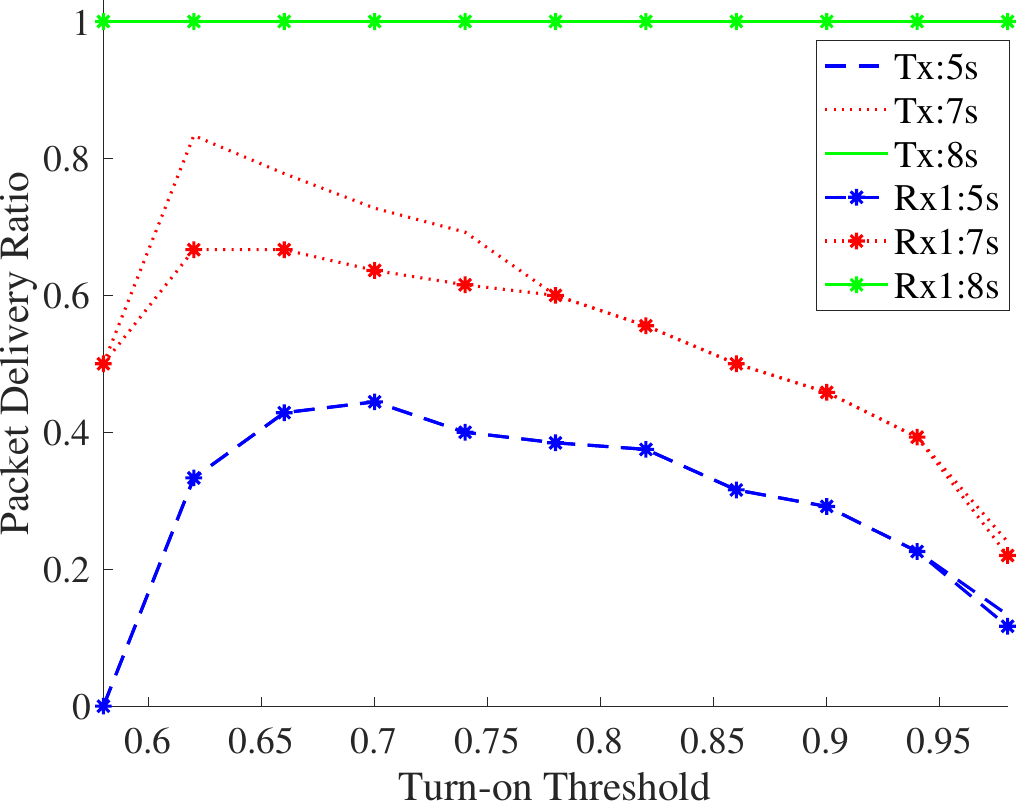}
			\label{fig:C1_DL1}}
		\subfigure[DL in RX window 2]{\includegraphics[width=0.65\columnwidth]{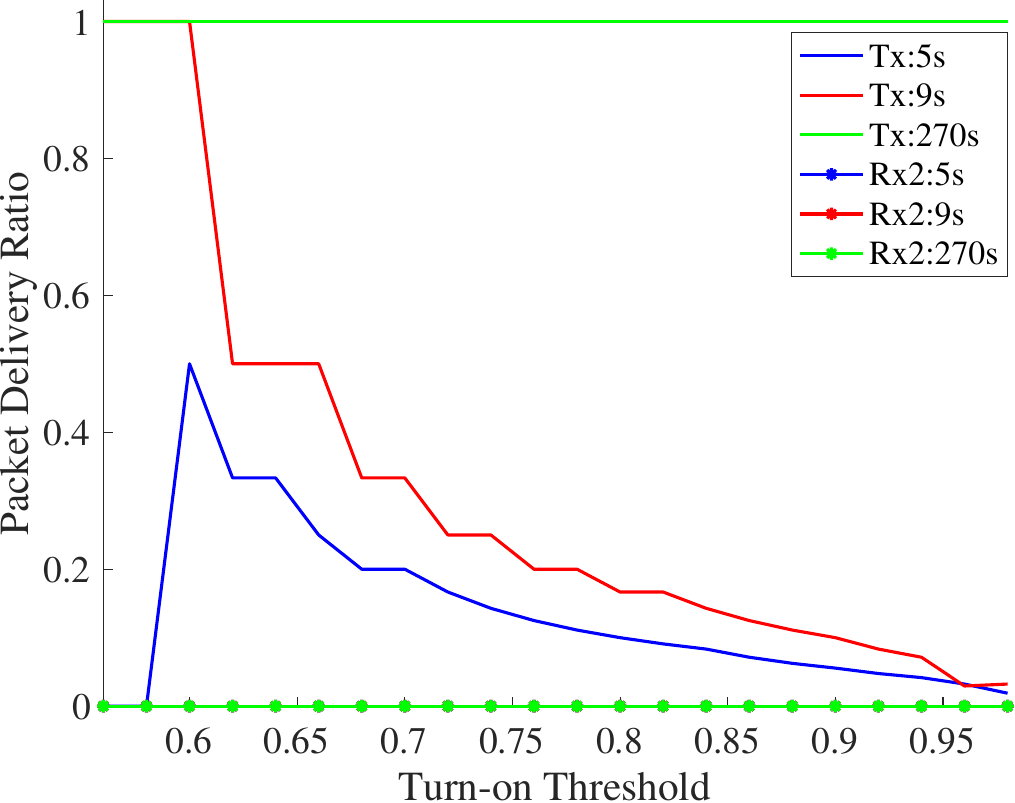}
			\label{fig:C1_DL2}}
		\caption{PDR for TX and RX when varying the turn-on threshold for a capacitance of $4.7 mF$ and different transmission intervals. Tx represents the PDR for transmissions, while Rx1/Rx2 represents the PDR of receiving it in the first or second window}
		\label{fig:C1_WuT}
	}
\end{figure*}
\begin{figure*}[t]
	\centering
	{
		\subfigure[No DL, only Tx PDR ]{\includegraphics[width=0.65\columnwidth]{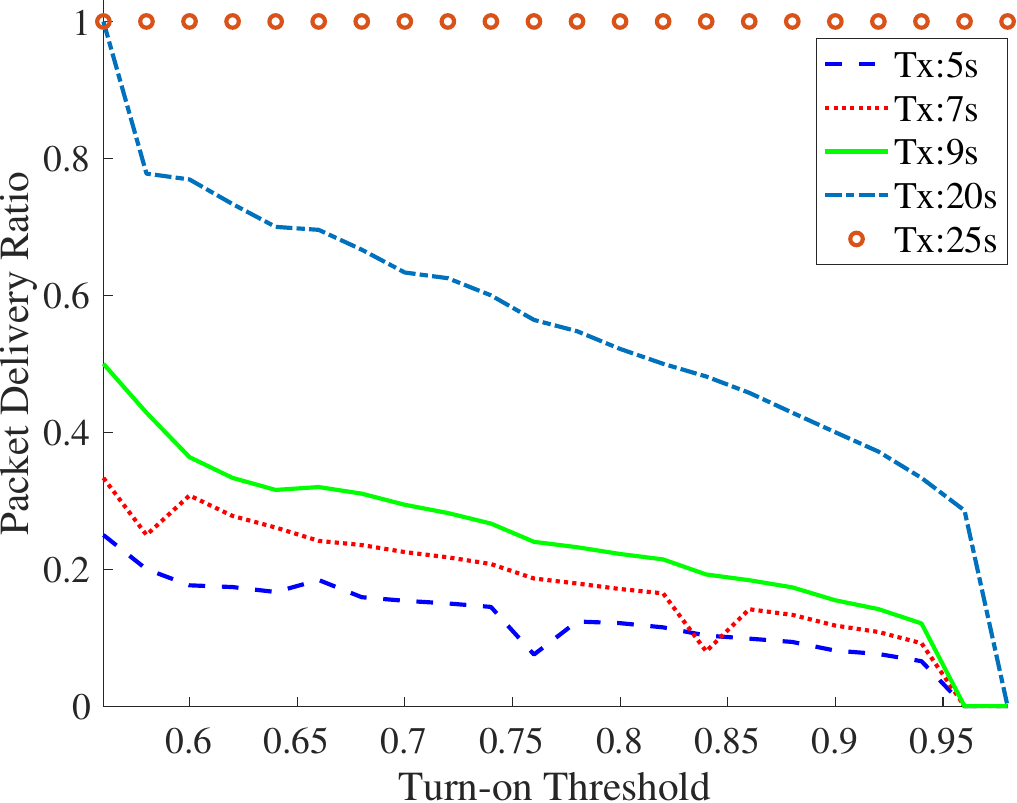}
			\label{fig:C2_NoDL}}
		\subfigure[DL in RX window 1]{\includegraphics[width=0.65\columnwidth]{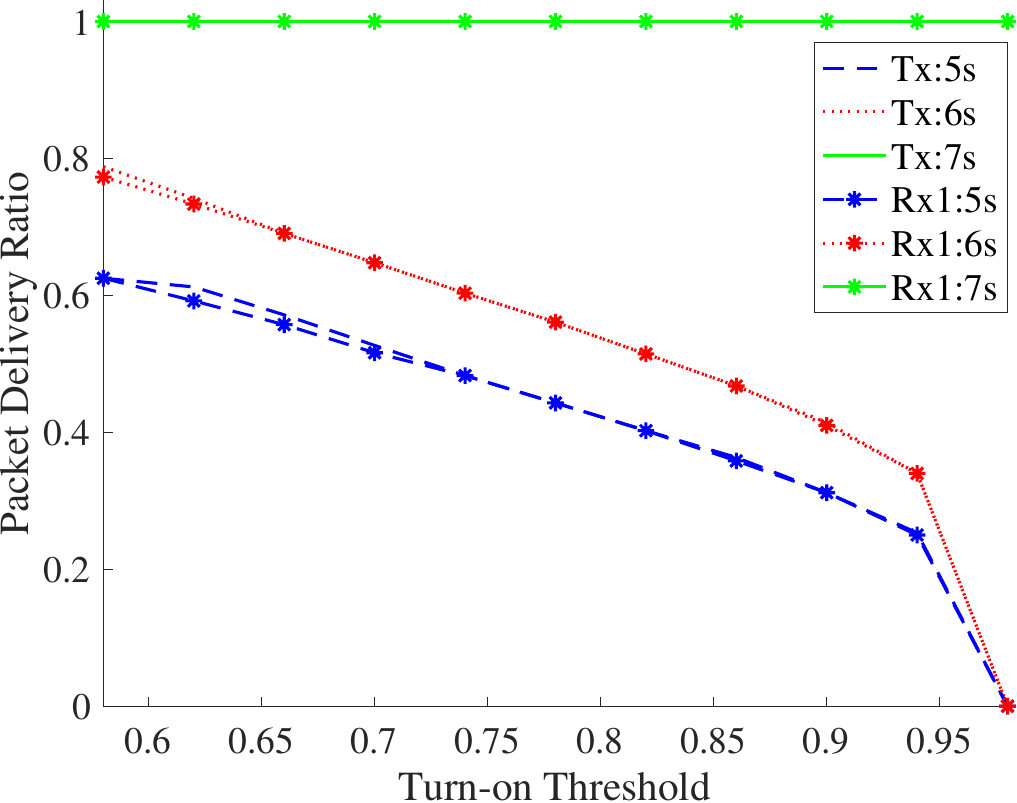}
			\label{fig:C2_DL1}}
		\subfigure[DL in RX window 2]{\includegraphics[width=0.65\columnwidth]{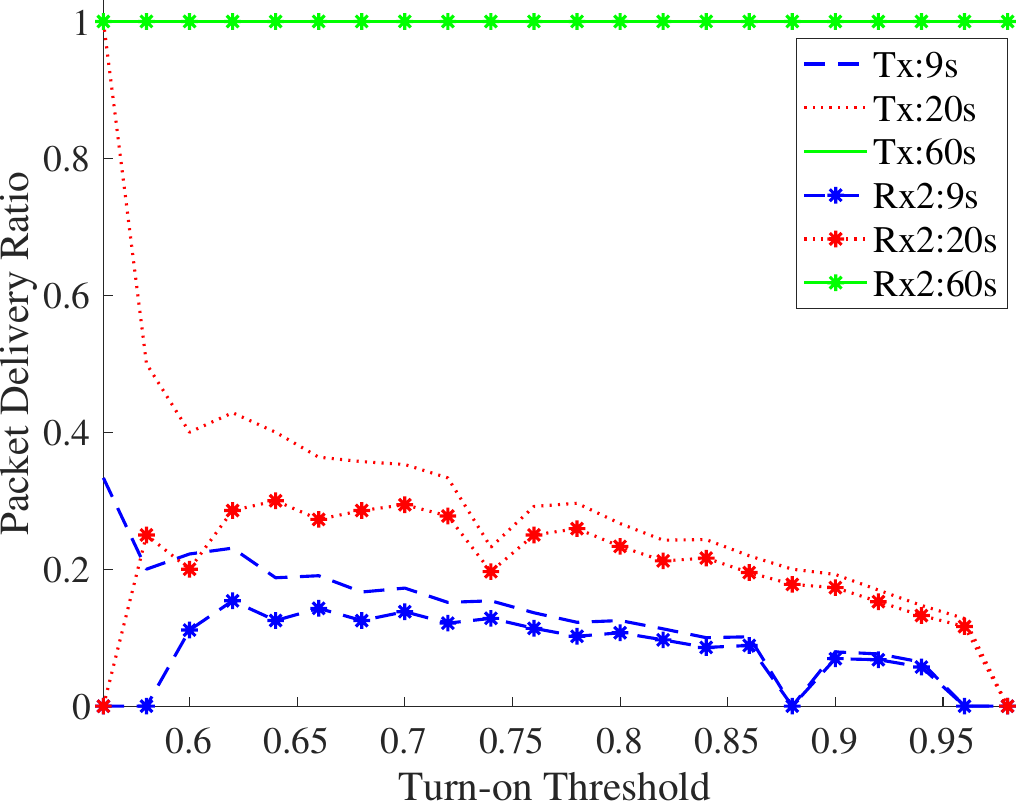}
			\label{fig:C2_DL2}}
		\caption{PDR for TX and RX when varying the turn-on threshold for a capacitance of $47 mF$ and different transmission intervals. Tx represents the PDR for transmissions, while Rx1/Rx2 represents the PDR of receiving it in the first or second window}
		\label{fig:C2_WuT}
	}
\end{figure*}

\subsubsection{Capacitance of $4.7 mF$}

Figure~\ref{fig:C1_WuT} shows the packet delivery ratio (PDR) as a function of the turn-on threshold and for different transmission intervals. These results consider a harvesting rate of $1mW$ (in line with indoor light harvesting). The Spreading Factor has been set to 7 and the UL and DL packet sizes are 16 and 1 Bytes respectively. Figure~\ref{fig:C1_NoDL} shows the case where $P_1=P_2=0$, i.e. no downlink is received, Figure~\ref{fig:C1_DL1} shows the case where $P_1=1$ and $P_2=0$ (the downlink is received in the first window) and Figure~\ref{fig:C1_DL2} shows the case where $P_1=0$ and $P_2=1$ (the downlink is received in the second window). 

As expected, packets can be transmitted more frequently when a small packet size is received in the first window (Figure~\ref{fig:C1_DL1}), where we can transmit every 8 seconds no matter what turn-on threshold is chosen. However, if no downlink is received (Figure~\ref{fig:C1_NoDL}), it is only possible to transmit every 9 seconds, where it is necessary to choose the right turn-on threshold (from 56\% up to 60\%). As it is also shown, the lowest turn-on threshold does not always provide the best PDR, when transmitting every 5 seconds, the turn-on threshold needs to be set to 60\% to be able to achieve a PDR of 0.5. If lower values are chosen, no packet will be able to be transmitted successfully.

For the worst case, where a DL is received in the second window, Figure~\ref{fig:C1_DL2} shows that this capacitance and energy harvesting rate do not allow the device to receive any DL in the second window. Besides that, since the device tries to receive it unsuccessfully, it wastes energy so successful transmissions will also become less frequent. 
Although choosing the proper turn-on threshold (lower than 60\%) will allow to transmit every 9 seconds, only transmission intervals of 270 seconds result in guaranteed success for all turn-on thresholds.

\begin{figure*}[t]
	\centering
	{
		\subfigure[No DL, only Tx PDR ]{\includegraphics[width=0.65\columnwidth]{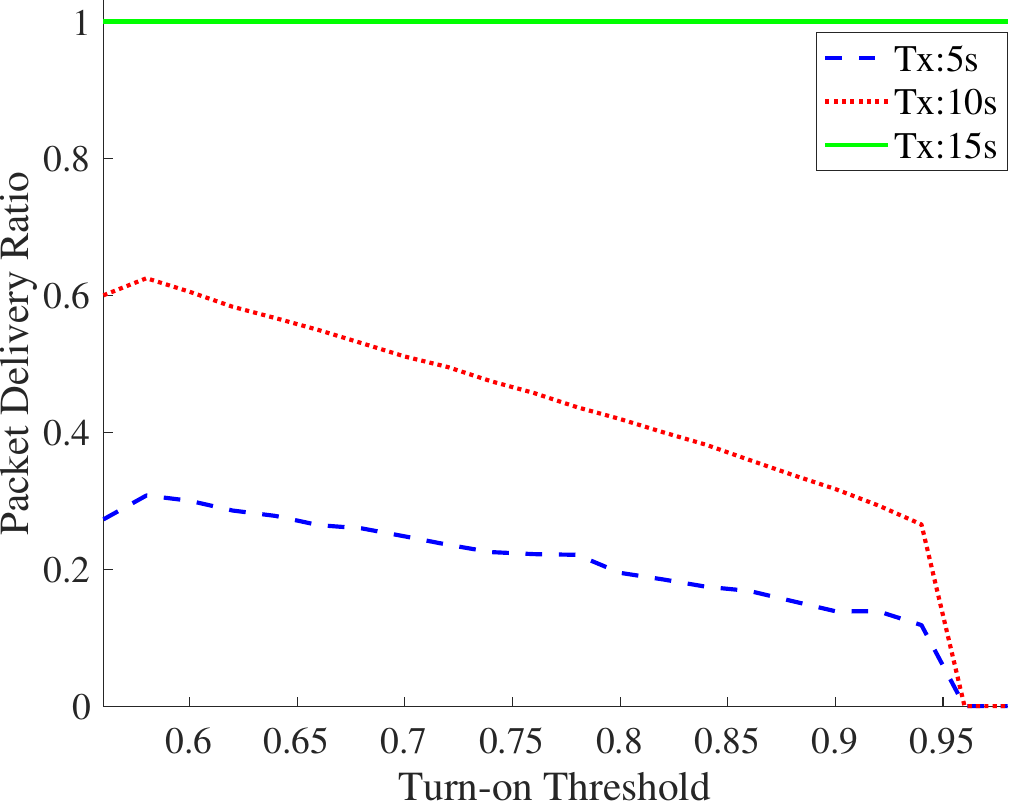}
			\label{fig:C3_NoDL}}
		\subfigure[DL in RX window 1]{\includegraphics[width=0.65\columnwidth]{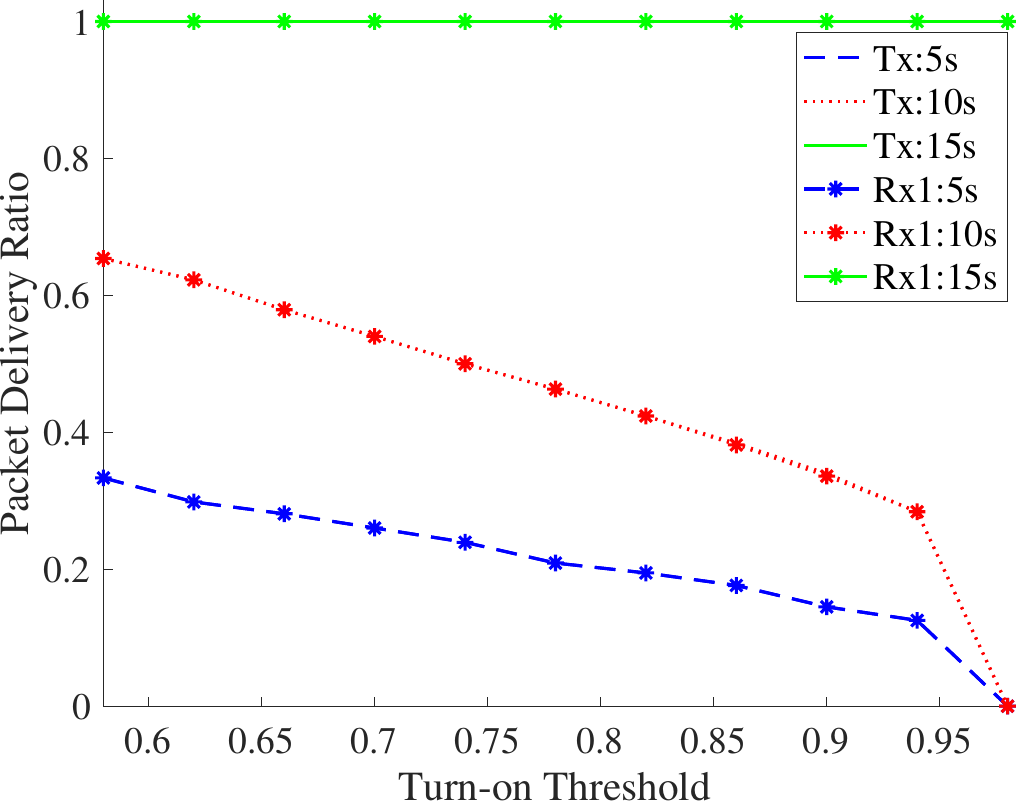}
			\label{fig:C3_DL1}}
		\subfigure[DL in RX window 2]{\includegraphics[width=0.65\columnwidth]{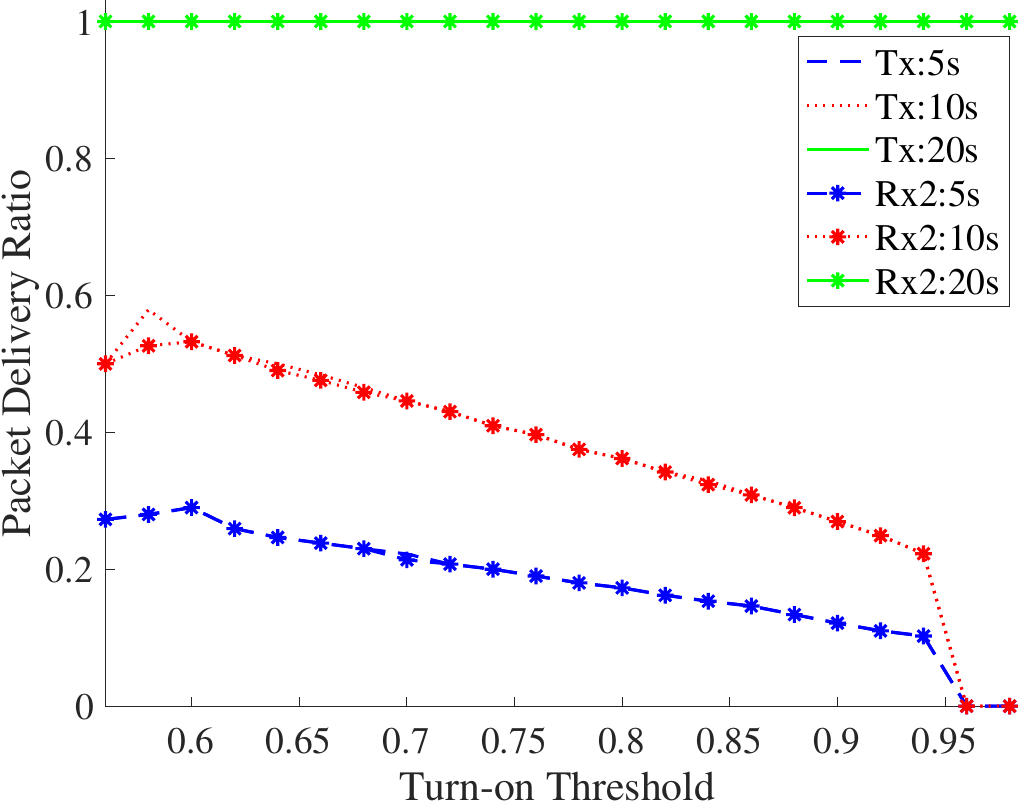}
			\label{fig:C3_DL2}}
		\caption{PDR for TX and RX when varying the turn-on threshold for a capacitance of $1 F$ and different transmission intervals. Tx represents the PDR for transmissions, while Rx1/Rx2 represents the PDR of receiving it in the first or second window}
		\label{fig:C3_WuT}
	}
\end{figure*}

\subsubsection{Capacitance of $47 mF$}
Figure~\ref{fig:C2_WuT}  shows the results when increasing the capacitance 10 times. In general, the  achievable transmission interval with $PDR=1$ is lower.
If no DL is received (Figure~\ref{fig:C2_NoDL}), only transmissions every 20 seconds guarantee a PDR of 1 if choosing the proper turn-on threshold (56\%). This is due to the fact that more time is needed to wake up bigger capacitors. However, if transmissions are done every 25 seconds or more, once the device has woken up, it does not turn-off since it never reaches the turn-off threshold and during the sleep periods it harvest enough energy to start a new cycle again after it. 

Figures~\ref{fig:C2_DL1} and \ref{fig:C2_DL2} show the case where DL is received in the first and in the second window respectively. When downlink is performed in the first window, transmissions in an interval of 7 seconds guarantee a PDR of 1,  for both uplink and downlink.
Figure~\ref{fig:C2_DL2} shows that increasing the capacitance allows to receive downlink packets in the second window even for low values of harvesting rates (1mW), for transmission intervals of 60 seconds.

\subsubsection{Capacitance of $1 F$}
In order to consider a supercapacitor, we have tested the behaviour of a capacitance of $1F$. We have considered more power-hungry scenarios. In this case, Figure~\ref{fig:C3_WuT} shows the results for transmissions of $SF11$ and uplink packet sizes of 48 Bytes. Downlink packet size is left at 1 Byte. For this case, we consider a value of $P_{harvester} = 10mW$, so similar transmission intervals can be achieved in these conditions. 
Figure~\ref{fig:C3_NoDL} shows that even for longer transmissions (high $SF$ and packet size) with normal harvesting rates ($P_{harvester} = 10mW$), low transmission intervals (15 seconds) can be achieved when using a larger capacitor. Additionally, Figure~\ref{fig:C3_DL1} shows that it is also possible to receive downlink packets for the same transmission intervals. When supercapacitors are used, the main problem is the time needed to charge them. Once they wake up, their energy is enough to perform UL and DL. As can be seen in Figure~\ref{fig:C3_DL2} a PDR of 1 is achieved for the second DL when the transmission interval is set to 20 seconds, for every possible turn-on threshold. As a downside, supercapacitors are much more expensive, support less charging cycles and are bulkier than normal capacitors.

\section{Conclusions}\label{sec:conclusions}

In this paper we have analyzed the feasibility of LoRaWAN Class A battery-less devices when considering both uplink and downlink transmissions. An analytical model of this system using a Markov Chain has been presented. The experimental validation of the model will be explored in the future. The accuracy of the Markov Chain, based on the $granularity$ parameter, 
has been evaluated by comparing its results with those obtained from a simulation model. When using a $granularity$ value of 750, which is computationally acceptable, we obtain an absolute error in the PDR lower than 0.003 for $90\%$ of the cases when using a turn-on threshold of $70\%$, and up to 0.02 for a turn-on threshold of $96\%$.

We have seen that LoRaWAN battery-less devices can be feasible for specific applications. We have analyzed the impact of the packet size and the transmission interval, and we have concluded that DL packet size highly affects the performance if the second reception window is used. For this reason, only small DL packet sizes should be consider in these devices (e.g., an ACK) and firmware updates are not possible. 

We have also shown that it is very important to properly design the device (capacitance and turn-on voltage threshold) depending on the application specification (transmission interval, packet size, downlink expectation) and environmental conditions (energy harvesting rate). Although bigger capacitances will allow more complex data transmissions (UL packet size up to 48 Bytes when using $SF11$), smaller capacitors would better fit when using the smallest $SF$ and small packet sizes (few Bytes), since they charge faster. Our results showed that a realistic capacitor of $4.7 mF$ can support $SF7$ for uplink and downlink at a low energy harvesting rate of 1mW, even when transmitting more than once per minute. $SF11$ can also be supported by a supercapacitor of $1F$ even if only 10mW of energy harvesting is provided.
It is also important to note that the turn-on voltage threshold influences the PDR, and depending on how often the application needs to send data and the size of the capacitor, it would be better to turn-on as soon as possible or wait to charge the needed amount of energy. But in general, the smallest turn-on threshold that allows the capacitor to charge up the energy level needed to complete a transmission-reception cycle, is the value that better performs in terms of PDR. For example, if a device needs to transmit every 10 seconds when using a capacitor of $4.7mF$ and no DL is received, it would perform better if the turn-on threshold is set to $60\%$ than a higher value.

To sum up, battery-less LoRaWAN Class A communications are feasible but the device turn-on voltage threshold significantly affects performance, so it will be necessary to dynamically tune it, based on environmental and application characteristics.

\section*{Acknowledgment}

Part of this research was funded by the Flemish FWO SBO S004017N IDEAL-IoT (Intelligent DEnse and Long range IoT networks) project and the FWO project G0B7915N “Modelling and Control of Energy Harvesting Wireless Sensor Networks”.
The computational resources and services used in this work were provided by the VSC (Flemish Supercomputer Center), funded by FWO and the Flemish Government - department EWI.

\ifCLASSOPTIONcaptionsoff
  \newpage
\fi



%
%

\bibliographystyle{IEEEtran}
\bibliography{./biblio}

%








\end{document}